\documentclass[11pt]{article}

\usepackage[english]{babel}
\usepackage[utf8]{inputenc}
\usepackage{amsmath}
\usepackage{graphicx}
\usepackage[colorinlistoftodos]{todonotes}
\usepackage{amssymb}
\usepackage{mathtools}
\usepackage{amsthm}
\usepackage[longnamesfirst]{natbib}
\usepackage{tikz}
\usepackage{bbm}
\usepackage{adjustbox}
\usepackage{etoolbox}
\usepackage{enumerate}
\usepackage{pdflscape}
\usepackage{adjustbox}
\usepackage{scrextend}
\usepackage{titling}
\usepackage{setspace}
\usepackage{bm}
\usepackage{mathrsfs}
\usepackage[margin=1.25in]{geometry}
\usepackage[normalem]{ulem}
\usepackage{hyperref}
\usepackage{cancel}
\hypersetup{pdfborder = 0 0 0,%
    pdftitle = A Justification of Conditional Confidence Intervals,%
    pdfauthor = Beutner Heinemann Smeekes,%
}

\renewcommand\qedsymbol{$\blacksquare$}


\newcommand\blfootnote[1]{%
  \begingroup
  \renewcommand\thefootnote{}\footnote{#1}%
  \addtocounter{footnote}{-1}%
  \endgroup
}

\usepackage{endnotes}

\numberwithin{equation}{section}

\newcommand{\EE}{\mathbb{E}}
\newcommand{\PP}{\mathbb{P}}
\newcommand{\Cov}{\mathbb{C}\mbox{ov}}

\newcommand{\R}{\mathbb{R}}
\newcommand{\N}{\mathbb{N}}
\newcommand{\iid}{\mbox{\textit{i.i.d.}}}

\newcommand{\X}{\mathbf{X}}
\newcommand{\Y}{\mathbf{Y}}
\newcommand{\x}{\mathbf{x}}

\theoremstyle{definition}
\newtheorem{assumption}{Assumption}
\newtheorem{assumption*}{Assumption}

\newtheorem{example}{Example}

\theoremstyle{plain}
\newtheorem{theorem}{Theorem}
\newtheorem{lemma}{Lemma}
\newtheorem{corollary}{Corollary}

\newtheorem{definition}{Definition}

\theoremstyle{remark}
\newtheorem{remark}{Remark}

\newenvironment{Example}[1][Example]{\begin{trivlist}
\item[\hskip \labelsep {\bfseries #1}]}{\end{trivlist}}

\title{A Justification of Conditional Confidence Intervals}

\author{Alexander Heinemann, Eric Beutner and Stephan Smeekes}

\date{\today}

\begin{document}

\begin{titlepage}
\begin{center}

\scshape
\LARGE{\textbf{A Justification of Conditional Confidence Intervals}}
\normalfont
\end{center}
\begin{center}
\par\vspace{0.5cm}%\vspace{1cm}
\text{Eric Beutner$^*$}
\hspace{1.5cm}
\text{Alexander Heinemann$^{**}$}
\hspace{1.4cm}
\text{Stephan Smeekes$^{***}$}
\par\vspace{0.5cm}%\vspace{1.5cm}
\text{Department of Quantitative Economics}
\par%\vspace{0.3cm}
\text{Maastricht University}
\par%\vspace{0.3cm}
\text{\today}
\vspace{1cm}
\end{center}

\blfootnote{\hspace{-0.6cm}
$^*$e-mail: \href{mailto:e.beutner@maastrichtuniversity.nl}{e.beutner@maastrichtuniversity.nl}\\
$^{**}$e-mail: \href{mailto:a.heinemann@maastrichtuniversity.nl}{a.heinemann@maastrichtuniversity.nl} (corresponding author)\\
$^{***}$e-mail: \href{mailto:s.smeekes@maastrichtuniversity.nl}{s.smeekes@maastrichtuniversity.nl}

The authors thank Franz Palm, Lenard Lieb, Denis de Crombrugghe, Hanno Reuvers and two anonymous referees for constructive comments and suggestions. This research was financially supported by the Netherlands Organisation for Scientific Research (NWO).
}

\begin{abstract}
To quantify uncertainty around point estimates of conditional objects such as conditional means or variances, parameter uncertainty has to be taken into account. Attempts to incorporate parameter uncertainty are typically based on the unrealistic assumption of observing two independent processes, where one is used for parameter estimation, and the other for conditioning upon. Such unrealistic foundation raises the question whether these intervals are theoretically justified in a realistic setting. This paper presents an asymptotic justification for this type of intervals that does not require such an unrealistic assumption, but relies on a sample-split approach instead. By showing that our sample-split intervals coincide asymptotically with the standard intervals, we provide a novel, and realistic, justification for confidence intervals of conditional objects. The analysis is carried out for a rich class of time series models.
\end{abstract}

\end{titlepage}
\newpage

\doublespacing

\section{Introduction}

One of the open questions in time series is how to quantify uncertainty around point estimates of conditional objects such as conditional means or conditional variances. A fundamental issue arises in the construction of confidence intervals that ought to capture the parameter estimation uncertainty contained in these objects. This fundamental issue stems from the fact that on one hand one must \textit{condition} on the sample as the past informs about the present, yet on the other hand one must allow the data up to now to be treated as \textit{random} to account for estimation uncertainty. 
The issue is well-recognized in the econometric literature, however in practice confidence intervals are commonly constructed by treating the sample \textit{simultaneously} as fixed and random. Frequently, such approach is motivated by presuming to have two independent processes. Assuming two independent processes with the same stochastic structure, using one for conditioning and one for the estimation of the parameters, bypasses the issue. It is a mathematically convenient assumption as in such case the uncertainty quantification reduces to an ordinary inferential problem. However, practitioners rarely have a replicate, independent of the original series, at hand with the exception of perhaps some experimental settings.  As such,  the intervals commonly constructed by practitioners lack a satisfactory theoretical justification. Therefore it is the objective of the present paper to develop a realistic justification for such confidence intervals around point estimates of conditional objects.

In the literature the fundamental issue described above is encountered in various ways. In the specific case of a first-order autoregressive (AR) process with Gaussian innovations, \cite{phillips1979sampling} investigates the statistical dependence between the ordinary least squares (OLS) estimator and the endogenous variable conditioned upon. He obtains an Edgeworth-type expansion for the distribution of the conditional mean and, further, studies forecasting, where the fundamental issue equally arises.\footnote{For prediction intervals some solutions have been discussed. We refer to Section \ref{sec:2.5}.} 
\citeauthor{lutkepohl2005new} (\citeyear{lutkepohl2005new}, p.\ 95) explicitly states a two-independent-processes assumption in connection with vector AR models. He postulates that such assumption is asymptotically equivalent to using only data not conditioned upon for estimation.
\cite{ing2003same} clearly distinguish between \textit{independent-realization} and \textit{same-realization} settings and study the (unconditional) mean-squared prediction error in the latter for an infinite-order AR process. In a companion paper they also provide a theoretical verification for order selection criteria for same-realization predictions and stress that it can be misleading to assume that the results for independent-realization settings carry over to those for corresponding same-realization cases \citep{ing2005order}. Other studies investigate parameter uncertainty by using resampling methods, that typically mimic a distribution in which the sample, or at least a subsample, is treated as fixed and random at the same time (cf.\ \citeauthor{pascual2004bootstrap}, \citeyear{pascual2004bootstrap}, \citeyear{pascual2006bootstrap}, \citeauthor{pan2016abootstrap}, \citeyear{pan2016abootstrap}, \citeyear{pan2016bbootstrap}). Aware of this paradox, \cite{kreiss2015discussion} points out that conditioning on observing specific \textit{in-sample} values affects the parameter estimator, but the effect is often erroneously disregarded. Deviating from the various bootstrap approaches, \cite{hansen2006interval} examines parameter uncertainty in interval forecasts in a classical statistical framework. Similar to a general regression framework, he conditions on an arbitrary fixed \textit{out-of-sample} value to avoid the issue. However, conditioning on arbitrary fixed out-of-sample values appears incompatible with the usual setup of dynamics in which we condition on the final value(s) of the sample.
Acknowledging the issue while avoiding the two independent processes argument bears careful statements as in \cite{francq2015risk} who write in view of this issue ``the delta method ... suggests" (p.\ 162). Similarly, \cite{pesaran2015time} notices that although such intervals ``have been discussed in the econometrics literature, the particular assumptions that underlie them are not fully recognized" (p.\ 389).

This paper provides a novel, and realistic, justification for commonly constructed confidence intervals around point estimates of conditional objects. Our solution is based on a simple sample-split approach and a weak dependence condition, which allows to partition our sample into two asymptotically independent subsamples. For a rich class of time series models we construct asymptotically valid sample-split intervals, without relying on the assumption of observing two independent processes, and show that these intervals coincide asymptotically with the intervals commonly constructed by practitioners. As will be argued below, an appropriate concept to study conditional confidence intervals is merging, a concept that generalizes weak convergence. To the best of our knowledge, except for \cite{belyaev2000weakly}, this paper is the only one to study merging in the context of conditional distributions. Moreover, our paper seems to be the first to employ merging of conditional distributions in time series. By employing this concept we avoid  unnatural assumptions such as observing $X_T=x$ (in dynamic models), losing the time index $T$, and instead explicitly acknowledge that the conditional objects vary over time.

The rest of the paper is organized as follows. Section \ref{sec:2.3} specifies the general setup and describes the argument of two independent processes as well as our sample-split  approach. In Section \ref{sec:2.4} we establish merging among the proposed and the two-independent-processes estimator  in probability under mild conditions. Further, we construct asymptotically valid sample-split intervals and show that these coincide asymptotically with the standard intervals. The extension to prediction is discussed in Section \ref{sec:2.5}. Section \ref{sec:2.6} concludes. The main proofs are collected in Appendix A, while Appendix B provides additional proofs of intermediate results.

\section{General Setup}
\label{sec:2.3}

\subsection{The General Prediction Function}
\label{sec:2.3.1}

Let $\{X_t\}$ be a real-valued stochastic process defined on the probability space $(\Omega,\mathcal{F},\PP)$. $\theta$ denotes a generic parameter vector of length $r \in \N$ and $\theta_0$ the true value, unknown to the researcher.\footnote{Generally, in particular throughout Section \ref{sec:2.3}, we do not distinguish between $\theta$ and $\theta_0$ if there is no cause for confusion. In Section \ref{sec:2.4} we explicitly use $\theta_0$ to avoid confusion.} Let $\Theta \subseteq \mathbb{R}^r$ be the corresponding parameter space.

Our general setup involves inference on an object that we call the \emph{prediction function}, which is a function of both the process $\{X_t\}$ and of the parameter $\theta$. It represents the random object of interest, and will typically express quantities such as a conditional mean or conditional variance (without conditioning on a specific value) as a function of the sample.
\begin{definition}\label{def prediction fct}
The \textit{prediction function} $\psi: \mathbb{R}^\infty \times \Theta \rightarrow \mathbb{R}$ is depending both on the parameter $\theta$ and the entire history of the process $\{X_t\}$, such that we can write the prediction of the quantity at time $T+1$, using data up to time $T$, as
\begin{equation} \label{eq:pr_func}
\psi_{T+1} :=
\psi (X_T, X_{T-1}, \ldots ; \theta).
\end{equation}
\end{definition}

With this setup we can describe most of the possible applications of interest. We now provide three examples to illustrate the prediction function.

\begin{example} \label{ex:2.1}
Suppose the time series $\{X_t\}$ follows an AR(1) process given by
\begin{align}
\label{eq:2.2.1}
X_t=\beta X_{t-1}+\varepsilon_t\:,
\end{align}
where $|\beta|<1$ and $\{\varepsilon_t\}$ are independent and identically distributed ($\iid$) with $\EE[\varepsilon_t]=0$. The conditional mean of $X_{T+1}$ given $X_T$ is given by
\begin{align}
\mu_{T+1} := \EE [X_{T+1} | X_T] = \beta X_T.
\end{align}
Using the prediction function we can then write $\mu_{T+1} = \psi(X_T, X_{T-1}, \ldots; \theta) = \beta X_T$ with $\theta = \beta$.
\end{example}

A more precarious example, due to its large popularity, is the conditional variance in a generalized autoregressive conditional heteroskedasticity (GARCH) model \citep{engle1982autoregressive,bollerslev1986generalized}. Whereas in the previous AR($1$) case it suffices to condition on the terminal observation, the subsequent Example 2 is more extreme as the entire sample contains information about the object of interest.
\begin{example} \label{ex:2.2}
Suppose $\{X_t\}$ follows a GARCH$(1,1)$ process given by $X_t = \sigma_t \varepsilon_{t}$ with
\begin{align}
\label{eq:2.2.5}
 \sigma_t^2 = \omega + \alpha X_{t-1}^2+ \beta \sigma_{t-1}^2\:,
\end{align}
where $\omega>0$, $\alpha \geq 0$, $1>\beta \geq 0$ and $\{\varepsilon_t\}$ are $\iid$ with $\EE[\varepsilon_t]=0$ and $\EE[\varepsilon_t^2]=1$. The model's recursive structure implies
\begin{align}
\label{eq:2.2.6}
 \sigma_{T+1}^2=& \frac{\omega}{1-\beta}+\alpha \sum_{k=0}^{\infty} \beta^k X_{T-k}^2.
\end{align}
It follows directly from \eqref{eq:2.2.6} that
\begin{equation*}
\sigma_{T+1}^2 = \psi (X_T, X_{T-1}, \ldots; \theta) =  \frac{\omega}{1-\beta}+\alpha \sum_{k=0}^{\infty} \beta^k X_{T-k}^2,
\end{equation*}
with $\theta = (\omega, \alpha, \beta)^\prime$, and $\Theta \subset(0,\infty)\times [0,\infty) \times [0,1)$.
\end{example}
The next example shows that for a large class of models the prediction function can be written in the form of Definition \ref{def prediction fct}.
\begin{example}
Following \cite{boussama2011stationarity}, consider a Markov chain of the form
\begin{align}
\label{eq:2.3.1}
S_t=\varphi(S_{t-1},X_t;\theta),
\qquad t=1,2,\dots
\end{align}
where $\varphi$ is some map $\varphi:\mathbb{R}^a\times \mathbb{R} \times \Theta \to \mathbb{R}^a$. Whereas $X_t$ is observable by the researcher at time $t$, $S_t$ may be unobservable or only partially observable. The object of interest $\psi_{T+1}$ is typically a function of the state of the Markov chain $S_T$, such that $\psi_{T+1} = \pi(S_T; \theta)$ for some function $\pi(\cdot)$. Through the recursion in \eqref{eq:2.3.1}, this is in turn a function of the past of $X_T$, such that we may write
\begin{equation*}
\psi_{T+1} = \pi(S_T; \theta) = \psi_{T+1} (X_T, X_{T-1}, \ldots; \theta).
\end{equation*}
Many stochastic processes are in fact Markov processes, including ARMA and GARCH models, several GARCH extensions such as \citeauthor{zakoian1994threshold}'s (\citeyear{zakoian1994threshold}) threshold GARCH, and the set of \textit{observation driven models} considered by \cite{blasques2016sample}. For further details we refer to \cite{beutner2017justification} and \cite{beutner2017technical}.
\end{example}

Note that in many cases, such as the GARCH(1,1) of Example \ref{ex:2.2}, the prediction function actually depends on the infinite past of the series. In order to express (an approximation of) the prediction function in terms of observable variables only, we would need to replace $X_t$ by $s_t$ for all  $t<1$, where $\{s_t\}$ is a sequence of (arbitrary) constants to which we refer as starting or initial values. For a fixed $T$, we accordingly define an approximate prediction function $\psi_{T+1}^s: \mathbb{R}^T \times \Theta \rightarrow \mathbb{R}$ that is only a function of observable variables as
\begin{equation} \label{eq:psi_s}
\psi_{T+1}^s (\X_{1:T}; \theta) := \psi (X_T, X_{T-1}, \ldots, X_1, s_0, s_{-1}, \ldots; \theta),
\end{equation}
where $\X_{1:T} = (X_1, \ldots, X_T)^\prime$. Note that, given the varying input of the left-hand side in (\ref{eq:psi_s}), we now actually have a sequence of (varying) functions for $T \in \mathbb{N}$.

In many cases the values far in the past are negligible for a wide range of values for $\{s_t\}$. Consequently, $\psi_{T+1}^s$ will be close to $\psi_{T+1}$. This property can be shown to hold for many different processes including the ones in the examples. We formalize the exact condition we need regarding the negligibility of the starting values in Assumption \ref{as:2.0}.b.

Although the prediction function typically represents a conditional object, we have not conditioned on anything yet in the definition. We therefore now extend the analysis by formally conditioning on observing a particular sample. Let $\x_{1:T} = (x_1,\dots,x_T)'$ denote a specific sample path of $\X_{1:T}$. Throughout the paper, we will discriminate between random variables and their realized counterparts by writing the former in capital and the latter in lowercase letters to avoid ambiguity. 

As we will consider sample splitting later on, we define notation that also allows for conditioning on only a subsample. For that purpose, let $t_1:t_2$ denote the (sub-)period from $t_1$ up to $t_2$, and correspondingly $\X_{t_1:t_2} = (X_{t_1}, \ldots, X_{t_2})^\prime$ for any integers $1 \leq t_1 \leq t_2 \leq T$, with a corresponding definition for the observed subsample $\x_{t_1: t_2}$. Furthermore, let $\X_{t_1:T}^{c} = (c_1, \ldots, c_{t_1-1}, X_{t_1}, \ldots, X_{T})^\prime$ denote the vector where all non-considered subsamples are replaced by a sequence of constants $\{c_t\}$, in a similar way as we did for the starting values. We can now formally define the conditional prediction function.

\begin{definition}\label{def prediction function conditional}
The \textit{prediction function conditional on observing $\X_{t_1:T} = \x_{t_1:T}$} is defined as
\begin{equation}
\psi_{T+1|t_1:T} := \psi_{T+1}^s (\x_{t_1:T}^c; \theta).
\end{equation}
\end{definition}
Note that we phrase the conditional prediction function directly in terms of the approximate prediction function $\psi_{T+1}^s$ rather than the true prediction function. We take this ``shortcut'' because we cannot observe $x_{0}, x_{-1}, \ldots$, so we cannot condition on those values anyway. Therefore, the ``true'' conditional object (which we might represent as $\psi_{T+1| -\infty: T}$), is, from an applied point of view, only the theoretical benchmark.

\begin{Example} \textbf{\ref{ex:2.1}.} \textit{(continued)} 
For the conditional mean of an AR(1) process, conditioning only on the terminal observation $X_T = x_T$ suffices; that is, for any $t_1 \geq 1$ and any sequence $\{c_t\}$, we have that
\begin{equation} \label{eq:psiAR}
\psi_{T+1|t_1:T} = \psi_{T+1}^s (\x_{t_1:T}^c; \theta) = \beta x_T = \psi_{T+1}^s (\x_{T:T}^c; \theta) = \psi_{T+1|T}.
\end{equation}
\end{Example}

\begin{Example} \textbf{\ref{ex:2.2}.} \textit{(continued)} 
For objects such as the conditional variance for the GARCH(1,1), the conditioning set and the sequence $\{c_t\}$ make a difference, as
\begin{equation}\label{eq psi T+1 | T GARCH(1,1)}
\begin{split}
\psi_{T+1|t_1:T} = \psi_{T+1}^s (\x_{t_1:T}^c; \theta) &=  \frac{\omega}{1-\beta} + \alpha \sum_{k=0}^{T-t_1} \beta^k x_{T-k}^2 \\
&\quad + \alpha \beta^{T-t_1} \sum_{k=1}^{t_1-1} \beta^k c_{t_1-k}^2 + \alpha \beta^T \sum_{k=0}^{\infty} \beta^k s_{-k}^2,
\end{split}
\end{equation}
which differs depending on the choice of $t_1$. However, as will be shown later, with an appropriate choice of $t_1$, our Assumption \ref{as:2.0}.b on the negligibility of the initial condition, also implies that the difference between $\psi_{T+1|t_1:T}$ and $\psi_{T+1|1:T}$ becomes negligible asymptotically.
\end{Example}

Before introducing estimators for $\theta$ let us discuss the objects we want to construct inference for. In principle there are two unknown objects one could develop statistical intervals for: $\psi_{T+1}^s (\X_{1:T}; \theta)$ (or slightly more generally $\psi_{T+1}^s (\X_{t_1:T}^c; \theta)$) and $\psi_{T+1}^s (\x_{t_1:T}^c; \theta).$ For a GARCH(1,1), for instance, the first would read as 
$$
\psi_{T+1}^s (\X_{1:T}; \theta) =  \frac{\omega}{1-\beta} + \alpha \sum_{k=0}^{T-1} \beta^k X_{T-k}^2 + \alpha \beta^T \sum_{k=0}^{\infty} \beta^k s_{-k}^2
$$
whereas the second with $t_1=1$ reads as 
$$
\psi_{T+1}^s (\x_{1:T}; \theta) =  \frac{\omega}{1-\beta} + \alpha \sum_{k=0}^{T-1} \beta^k x_{T-k}^2 + \alpha \beta^T \sum_{k=0}^{\infty} \beta^k s_{-k}^2
$$
or more generally, if $t_1$ is not taken to be equal to one, as in (\ref{eq psi T+1 | T GARCH(1,1)}). While statistical intervals for both objects can be constructed we focus here on conditional inference, i.e.~on intervals for $\psi_{T+1|t_1:T}=\psi_{T+1}^s (\x_{t_1:T}^c; \theta)$. In a time series context intervals for $\psi_{T+1|t_1:T}$ are motivated by the \textit{relevance property} of \cite{kabaila1999relevance} which postulates that intervals should relate to what actually happened during the sample period opposed to what might have happened. Indeed, intervals for $\psi_{T+1|t_1:T}$ can theoretically be shown to be considerably shorter than the intervals for their unconditional counterparts. While the unconditional objects might lead to conceptually easier analysis, our focus on the conditional objects is therefore not only theoretically but also empirically relevant.

\subsubsection{Estimating the Prediction Function}
As $\theta$ is unobserved, we need to estimate it. We assume that the estimator is based on a subsample $1:T_E$ (with $1 \leq T_E \leq T$) of the process $\{X_t^{E}\}$ which is potentially a \emph{different} sample than $\{X_t\}$ that arises in the prediction function. The estimator of $\theta$ based on $\X_{1:T_E}^E=(X_1^E,\dots, X_{T_E}^E)^\prime$ will be denoted by $\hat{\theta} (\X_{1:T_E}^E)$. The introduction of $\{X_t^{E}\}$  serves three purposes: first, using a different process allows us to formulate the two-independent-processes argument where $X_t^E = Y_t$, with $\{Y_t\}$ independent of $\{X_t\}$, $T_E = T$ and an interval is constructed for $\psi_{T+1|1:T}$. Second, it will allow us to discuss the standard approach where $ X_t^E = X_t$,  $T_E=T$, and an interval is constructed for $\psi_{T+1|1:T}$. Please note already here that this means that the same variables that arise in the prediction function are also used for estimating $\theta$. Third, it allows us to define the sample splitting approach which we study here. In this approach $ X_t^E = X_t$ and an interval is constructed for $\psi_{T+1|T_P:T}$ with $T_E < T_P$ (with $1 < T_P \leq T$) such that in contrast to the standard approach different subsamples are used for prediction and estimation.  \\
Before we illustrate why the standard approach is problematic for constructing and evaluating conditional intervals, we need to define the final building block of prediction function estimation: the conditional prediction function estimator.

\begin{definition}\label{def prediction function estimator conditional on observing}
Let $1 \leq T_P \leq T$. Define the \textit{prediction function estimator conditional on observing $\X_{T_P:T} = \x_{T_P:T}$} as
\begin{equation}
\widehat{\psi}_{T+1|T_P:T} := \widehat{\psi}_{T+1} (\x_{T_P:T}^{c}, \X_{1:T_E}^E) = \psi_{T+1}^s (\x_{T_P:T}^c, \hat{\theta} (\X_{1:T_E}^E)).
\end{equation}
\end{definition}
Note that in the above definition we do not condition on the sample  $\X_{1:T_E}^E = \x_{1:T_E}^E$ that is used to estimate $\theta$. The reason for not conditioning on $\X_{1:T_E}^E = \x_{1:T_E}^E$ is that the goal is to preserve the randomness in the second argument of $\psi_{T+1}^s$, i.e.~in $\hat{\theta} (\X_{1:T_E}^E)$, and consequently in $\widehat{\psi}_{T+1|T_P:T}$. Hence, if this goal is achieved we can use the (non-degenerate) conditional (on $\X_{T_P:T} = \x_{T_P:T}$) distribution of $\widehat{\psi}_{T+1|T_P:T}$ to construct confidence intervals for $\psi_{T+1|T_P:T}$. Having this said let us have a closer look at the standard approach. As mentioned above in the standard approach one has $X_t^E = X_t$, $T_P=1$ and $T_E=T$. Hence, denoting by $\widehat{\psi}_{T+1|1:T}^{STA}$ the ``standard'' estimator of the prediction function conditional on observing $\X_{1:T} = \x_{1:T}$, it becomes
\begin{equation} \label{eq:STA}
\widehat{\psi}_{T+1|1:T}^{STA}  :=\hat{\psi}_{T+1}(\x_{1:T},\x_{1:T})=\psi_{T+1}^s (\x_{1:T}, \hat{\theta} (\x_{1:T})).
\end{equation} 
Notice that there is no capital $\X$ in (\ref{eq:STA}) because there is only one sample and one typically conditions on all values of this sample. Hence,  (\ref{eq:STA}) is non-random and thus does not have a distribution that could be used to construct intervals. Instead, to still be able to construct a ``standard-looking'' interval in practice, researchers typically implicitly rely on the (approximate) quantiles of the estimator
\begin{align}
\label{eq:2.3.7}
\hat{\psi}_{T+1|1:T}^{STA*} := \hat{\psi}_{T+1}(\x_{1:T},\X_{1:T})=\psi_{T+1}^s(\x_{1:T}; \hat{\theta}(\X_{1:T})).
\end{align}
It is well understood in the literature that considering the sample as random and non-random at the same time as in (\ref{eq:2.3.7}) does not provide a fully satisfactory justification of the intervals used in practice. For the readers not so familiar with the problem just discussed we provide two examples that both illustrate the problem arising from (\ref{eq:2.3.7}). The examples illustrate that the severity of the problem may vary; ranging from only complicating the analysis (Example \ref{ex:2.1}) to making the analysis impossible (Example \ref{ex:2.2}).
\begin{Example} \textbf{\ref{ex:2.1}.} \textit{(continued)} 
For the AR(1), we know from (\ref{eq:psiAR}) that $\psi_{T+1|1:T}=\psi^s_{T+1}(\mathbf{x}_{1:T},\theta)=\beta x_T$. Estimating $\beta$ by OLS, say $\hat{\beta}(\X_{1:T})$,
the estimator in (\ref{eq:2.3.7}) becomes
\begin{equation}\label{eq sta* ar 1}
\hat{\psi}_{T+1|1:T}^{STA*} = \psi_{T+1} (\x_{1:T}; \hat{\theta}(\X_{1:T})) = \psi_{T+1}^s (\x_{1:T}, \hat{\beta} (\X_{1:T})) = \hat{\beta} (\X_{1:T}) x_T.
\end{equation}
Note the discrepancy in treating the terminal observation as random in the estimation sample, yet fixed for the prediction sample. To construct an interval for $\beta x_T$, one uses 
\begin{align}\label{eq:2.2.3}
\sqrt{T}\big(\hat{\beta}(\mathbf{X}_{1:T})-\beta\big)\overset{d}{\to} N(0,\sigma_\beta^2)
\end{align}
with $\sigma_\beta^2=1-\beta^2$ (cf.\ \citeauthor{hamilton1994time}, \citeyear{hamilton1994time}, p.\ 215) and that one can estimate the variance of this normal distribution by $\hat{\sigma}_\beta^2(\mathbf{X}_{1:T})=1-\hat{\beta}(\mathbf{X}_{1:T})^2$. Then an interval for $\beta x_T$ is typically constructed the following way:
\begin{align}
\label{eq:2.2.4}
\hat{\beta}(\mathbf{X}_{1:T})x_T\pm \Phi^{-1}(\gamma/2)\:x_T\:\hat{\sigma}_\beta(\mathbf{X}_{1:T})/\sqrt{T}\:,
\end{align}
where $\Phi^{-1}$ denotes the standard normal quantile function. However, the interval in \eqref{eq:2.2.4} is hard to interpret as the terminal observation is treated simultaneously as fixed and random. In essence, researchers typically approximate the distribution of  $\sqrt{T}(\hat{\beta}(\mathbf{X}_{1:T})-\beta) x_T$ instead of the conditional distribution of $\sqrt{T}(\hat{\beta}(\mathbf{X}_{1:T})-\beta) X_T$ given $X_T=x_T$. The approximation of the latter appears rather cumbersome because even the rather simple condition $X_T=x_T$ has an influence on the whole series $\mathbf{X}_{1:T}$ (\citeauthor{kreiss2015discussion}, \citeyear{kreiss2015discussion}). Despite the challenge, \cite{phillips1979sampling} obtains such approximation based on Edgeworth expansions in the case of $\varepsilon_t\overset{iid}{\sim}N\big(0,\sigma^2_\varepsilon\big)$.
\end{Example}

\begin{Example} \textbf{\ref{ex:2.2}.} \textit{(continued)} 
For the conditional variance of the GARCH(1,1), the standard estimator of the prediction function conditional on  $\X_{1:T} = \x_{1:T}$, is given by
\begin{align}
\label{eq:2.2.9}
\begin{split}
\hat{\sigma}_{T+1|1:T}^{2\;STA} = \psi_{T+1}^s \big(\mathbf{x}_{1:T};\hat{\theta}(\mathbf{x}_{1:T})\big)
&= \frac{\hat{\omega}(\mathbf{x}_{1:T})}{1-\hat{\beta}(\mathbf{x}_{1:T})}+\hat{\alpha}(\mathbf{x}_{1:T}) \sum_{k=0}^{T-1} \hat{\beta}(\mathbf{x}_{1:T})^k x_{T-k}^2\\
&\quad +\hat{\alpha}(\mathbf{x}_{1:T}) \hat{\beta}(\mathbf{x}_{1:T})^T \sum_{k=0}^{\infty} \hat{\beta}(\mathbf{x}_{1:T})^k s_{-k}^2 \:,
\end{split}
\end{align}
where $\hat{\theta}(\mathbf{x}_{1:T})=\big(\hat{\omega}(\mathbf{x}_{1:T}),\hat{\alpha}(\mathbf{x}_{1:T}),\hat{\beta}(\mathbf{x}_{1:T})\big)'$ is some estimate for $\theta$ depending on $\mathbf{x}_{1:T}$. Clearly, (\ref{eq:2.2.9}) illustrates for the GARCH(1,1) the above mentioned problem that the standard estimator is not random (after conditioning). For the GARCH(1,1) the estimator in (\ref{eq:2.3.7}) whose quantiles are used for an interval reads as 
\begin{align}
\label{eq:2.2.10}
\begin{split}
\hat{\sigma}_{T+1}^{2\;STA*}=\psi_{T+1}\big(\mathbf{x}_{1:T};\hat{\theta}(\mathbf{X}_{1:T})\big)
&= \frac{\hat{\omega}(\mathbf{X}_{1:T})}{1-\hat{\beta}(\mathbf{X}_{1:T})}+\hat{\alpha}(\mathbf{X}_{1:T}) \sum_{k=0}^{T-1} \hat{\beta}(\mathbf{X}_{1:T})^k x_{T-k}^2\\
&\quad +\hat{\alpha}(\mathbf{X}_{1:T}) \hat{\beta}(\mathbf{X}_{1:T})^T \sum_{k=0}^{\infty} \hat{\beta}(\mathbf{X}_{1:T})^k s_{-k}^2 \:.
\end{split}
\end{align}
This quantity exemplifies for the GARCH(1,1) that the complete sample is regarded as random and non-random at the same time. While for the AR(1) this complicated the analysis, yet not made it impossible, the dependence on the complete sample here makes it difficult to use this quantity to make meaningful probabilistic statements.
\end{Example}

\subsection{Argument of Two Independent Processes} \label{sec:2.3.2}
The argument of two independent processes can at least be traced back to \cite{akaike1969fitting}, who studies the prediction of AR time series. It reoccurs in \citeauthor{lewis1985prediction} (\citeyear{lewis1985prediction}, p.\ 394):
``...the series used for estimation of parameters and the series used for prediction are generated from two independent processes which have the same stochastic structure." The same argument also appears in \citeauthor{lutkepohl2005new} (\citeyear{lutkepohl2005new}, p.\ 95) and in \cite{dufour2010short}. Let $\{Y_t\}$ be a process independent of $\{X_t\}$ defined on the same probability space $(\Omega,\mathcal{F},\PP)$ with $\{Y_t\}$ having the same stochastic structure as $\{X_t\}$. In addition to the sample $\X_{1:T}$ of the process $\{X_t\}$, suppose there is a sample $\Y_{1:T} = (Y_1,\dots,Y_T)'$ of the process $\{Y_t\}$ that we use as estimation sample, that is $\X_{1:T}^E = \Y_{1:T}$. In this situation we denote the conditional prediction function estimator of Definition \ref{def prediction function estimator conditional on observing} by $\widehat{\psi}_{T+1|1:T}^{2IP}$ and it equals
\begin{align}
\label{eq:2.3.11}
\widehat{\psi}_{T+1|1:T}^{2IP} := \widehat{\psi}_{T+1} (\x_{1:T}, \Y_{1:T})= \psi_{T+1}^s (\x_{1:T}; \hat{\theta} (\Y_{1:T})).
\end{align}
Notice that (\ref{eq:2.3.11}) does not have the same shortcoming as (\ref{eq:2.3.7}) because even if we consider $\x_{1:T}$ to be known we can nevertheless consider $\hat{\theta}(\Y_{1:T})$ to be random and can hence use its distribution to construct intervals. Throughout this paper, we call \eqref{eq:2.3.11}, the 2IP (two independent processes) estimator. Then, a conditional interval $I_\gamma^{2IP}(\x_{1:T}, \Y_{1:T})$ can be based on the (approximate) quantiles of $\widehat{\psi}_{T+1|1:T}^{2IP}$.
It satisfies
\begin{align}
\label{eq:2.3.13}
\PP \Big[ I_\gamma^{2IP}(\x_{1:T}, \Y_{1:T}) \ni \psi_{T+1|1:T}\Big|\X_{1:T} = \x_{1:T} \Big]
\underset{(\approx)}{=} 1 -\gamma,
\end{align}
with the approximate sign indicating asymptotic equivalence. Note that the independence implies that the distribution of $\Y_{1:T}$ in (\ref{eq:2.3.13}) does not depend on the realization $\x_{1:T}$, yet the statement does depend on $\x_{1:T}$ because the interval depends on it (for the AR(1) this can be directly seen from (\ref{eq:2.2.4}) when replacing $\X_{1:T}$ by $\Y_{1:T}$). Although the 2IP approach is statistically sound, it assumes two independent processes with the same stochastic structure. \cite{phillips1979sampling} points out that the assumption ``is quite unrealistic in practical situations" (p.\ 241). Indeed, it is difficult to imagine this assumption to be satisfied in any real-life application beyond experimental settings. Moreover, as only one sample realization is available, to compute the estimate of the interval $I_\gamma^{2IP}(\x_{1:T}, \Y_{1:T})$ it is frequently suggested to take $ \Y_{1:T} = \x_{1:T}$, violating the independence assumption. Thus, the 2IP approach appears to be a rather questionable justification for the usual interval, and as such, in this paper we provide an alternative, realistic, justification of  (asymptotically equivalent) intervals based on sample splitting.

\subsection{Sample-split Estimation}
\label{sec:2.3.3}

An intuitive motivation for the sample-split approach is the successive decline of the influence of past observations present in a substantial class of time series models. This property permits to split our sample into two (asymptotically) independent subsamples. Consider the end point of the estimation sample, $T_E$, and the starting point of the prediction sample, $T_P$ satisfying $1< T_E < T_P \leq T$, such that the two samples are non-overlapping. In this situation we denote the conditional prediction function estimator of Definition \ref{def prediction function estimator conditional on observing} by $\widehat{\psi}_{T+1|T_P:T}^{SPL}$ and it is given by
\begin{align}
\label{eq:2.3.16}
\widehat{\psi}_{T+1|T_P:T}^{SPL} := \widehat{\psi}_{T+1} (\x_{T_P:T}^c, \X_{1:T_E}) = \psi_{T+1}^s (\x_{T_P:T}^c; \hat{\theta} (\X_{1:T_E})).
\end{align}
Throughout the paper, we call (\ref{eq:2.3.16}) the SPL estimator (due to SPLitting). Similar to the two sample approach, we can consider the first argument of $\psi_{T+1}^s$ in (\ref{eq:2.3.16}) as given and the second argument as random since the subsamples are non-overlapping. A conditional interval $I_\gamma^{SPL}(\x_{T_P:T}, \X_{1:T_E})$ can be constructed such that
\begin{align}
\label{eq:2.3.17}
&\PP \Big[ I_\gamma^{SPL}(\x_{T_P:T}, \X_{1:T_E}) \ni \psi_{T+1|T_P:T} \Big|\X_{T_P:T} = \x_{T_P:T} \Big]\underset{(\approx)}{=} 1-\gamma\:.
\end{align}
This statement does make sense as there is still randomness in $\hat{\theta}(\X_{1:T_E})$ since $\X_{1:T_E}$ is not conditioned upon, yet the last $T - T_P+1$ values of $\{X_t\}_{t=1}^T$ are fixed such that their randomness is not taken into account. Similar to (\ref{eq:2.3.13}), the statement in (\ref{eq:2.3.17}) does depend on $\x_{T_P:T}$ and in contrast to $\mathbf{x}_{1:T}$ in (\ref{eq:2.3.13}) the realization $\x_{T_P:T}$ may influence the distribution of $\X_{1:T_E}$. However, as said at the beginning of this subsection the idea of the sample split approach is that this dependence will vanish asymptotically.
\begin{remark}
\label{rem:2.1}
In Section \ref{sec:2.4} we will discuss how $T_E$ and $T_P$ should be chosen from an asymptotic perspective to ensure that our regularity conditions are fulfilled. As we only consider sample splitting as a theoretical approach to validate commonly constructed conditional confidence intervals, these asymptotic guidelines are sufficient for our purposes and we do not have to consider how to choose $T_E$ and $T_P$ in practice. Of course, one could use the sample-split approach in practice to construct confidence intervals. While one would gain (near) independence between the two subsamples, this would come at a cost of estimation precision as fewer observations are used for parameter estimation. For the Gaussian AR($1$) setting, \cite{phillips1979sampling} derives asymptotic expansions for the case where, in our notation, $T_E = T-l$ and $T_P = T$ for some $l \geq 0$, showing that even in this simple case there is indeed a trade-off as described above and the optimal choice of $l$ is unclear. An interesting extension of our analysis would therefore be to investigate the optimal choices of $T_E$ and $T_P$ to achieve the most accurate confidence intervals in small samples. However, this choice is likely to be highly dependent on the specific model and as such would have to be investigated on a case-by-case basis. This is therefore outside the scope of the current paper.
\end{remark}

\section{Asymptotic Justification}
\label{sec:2.4}

In this section, we connect the sample-split procedure of Section \ref{sec:2.3.3} with the two-independent-samples approach of Section \ref{sec:2.3.2}. First, in Section \ref{sec:2.4.1}, we show that the notion of weak convergence is inadequate to study asymptotic closeness for objects that vary over time and discuss the concept of merging. Then, in Section \ref{sec:2.4.2} we link the 2IP and the SPL estimator by proving that their conditional distributions merge in probability (Theorem \ref{thm:2.1}). Thereafter, in Section \ref{sec:2.4.3}, we construct asymptotically valid intervals (Theorem \ref{thm:2.2}) and show that the sample-split intervals coincide asymptotically with the  intervals commonly constructed by practitioners (Theorem \ref{thm:2.3}). Last, in Section \ref{sec:2.4.4}, we state intervals of reduced form and simplified theoretical results under asymptotic normality of the parameter estimator.

\subsection{Merging}\label{sec:2.4.1}

To illustrate the inappropriateness of weak convergence in the context considered here, we revisit Example \ref{ex:2.1} for the 2IP approach and the SPL approach, which shows that studying asymptotic closeness between conditional distributions is often complicated by the absence of a limiting distribution.

\begin{Example} \textbf{\ref{ex:2.1}.} \textit{(continued)}
For the 2IP approach \eqref{eq:2.2.3} implies that $\sqrt{T}\big(\hat{\beta}(\mathbf{Y}_{1:T})-\beta\big)\overset{d}{\to} N(0,\sigma_\beta^2)$
and it entails that $\sqrt{T}(\hat{\beta}(\mathbf{Y}_{1:T})-\beta)x$ converges weakly to $N(0,\sigma_\beta^2x^2)$ for any fixed $x\neq 0$. Further, the result suggests that the conditional distribution of $\sqrt{T}(\hat{\beta}(\mathbf{Y}_{1:T})-\beta)X_T$ given $X_T=x_T$, which is just the distribution of $\sqrt{T}(\hat{\beta}(\mathbf{Y}_{1:T})-\beta)x_T$, is asymptotically close to $N(0,\sigma_\beta^2x_T^2)$. Similarly, for the SPL-approach with $T_E \slash T \rightarrow 1$ we have  $\sqrt{T}\big(\hat{\beta}(\mathbf{X}_{1:T_E})-\beta\big)\overset{d}{\to} N(0,\sigma_\beta^2)$ which suggests as well (if the gap between $T_E$ and $T$ is large enough which will be  formally specified below) that the conditional distribution of $\sqrt{T_E}(\hat{\beta}(\mathbf{X}_{1:T_E})-\beta)X_T$ given $X_T=x_T$ is also close to $N(0,\sigma_\beta^2x_T^2)$. For both approaches the approximating distribution $N(0,\sigma_\beta^2x_T^2)$ varies with $T$ through the terminal realization $x_T$. Note that the concept of weak convergence is not applicable in this context to characterize this asymptotic closeness, as it requires a (fixed) limiting distribution, which is absent here.
\end{Example}
Next, we discuss what closeness means in the absence of a limiting distribution. To do so, first recall that weak convergence of a sequence of cdfs $\{F_T\}$ on $\mathbb{R}^k$ with $k \in \N$, i.e.\  $F_T(x)\to F(x)$ for all continuity points of $F$, can alternatively be defined by $d_{BL}(F_T,F) \rightarrow 0$. Here $d_{BL}$ denotes the bounded Lipschitz metric defined by
\begin{align}
\label{eq:2.4.1}
d_{BL}(F,G)=\sup \bigg\{ \bigg|\int f d(F-G)\bigg|:||f||_{BL}\leq 1\bigg\}\:,
\end{align}
where for any real-valued function $f$ on $\R^k$ one puts $||f||_{BL}=\sup_x \big|f(x)\big|+\sup_{x \neq y}\frac{|f(x)-f(y)|}{||x-y||}$, with $||\cdot||$ denoting the Euclidean norm, i.e.\ $||A||=\sqrt{tr(A'A)}$ for any vector or matrix $A$. Following \cite{dudley2002real} (see \cite{d1988merging} and \cite{davydov2009asymptotic} for related work) we state the following definition.
\begin{definition} \textit{(Merging)}
\label{def:2.1}
Two sequences of cdfs $\{F_T\}$ and $\{G_T\}$ are said to \textit{merge} if and only if  $d_{BL}(F_T,G_T) \to 0$ as $T\to \infty$.
\end{definition}
Note that weak convergence can be seen as a special case of merging with $G_T=G$ for all $T \in \N$.\\
While merging is appropriate to capture the asymptotic closeness of the conditional distribution of $\sqrt{T}(\hat{\beta}(\mathbf{Y}_{1:T})-\beta)X_T$ and $N(0,\sigma_\beta^2x_T^2)$ for a given sample $X_T=x_T$, we now extend the concept in a way that allows us to deal with asymptotic closeness when we do not condition on a particular sample. The necessity of this definition can again be exemplified by the AR(1), which also illustrates how we will deal the dependence of the statements in  (\ref{eq:2.3.13}) and in (\ref{eq:2.3.17}) on the sample that we mentioned below these equations. For instance, in Example \ref{ex:2.1} as described at the beginning of this section, the goal would be to formalize a statement like `when $T$ is large, the probability of all $x_T$ such that the distribution of $\sqrt{T}(\hat{\beta}(\mathbf{X}_{1:T})-\beta)x_T$ merges with that of $\sqrt{T_E}(\hat{\beta}(\mathbf{X}_{1:T_E})-\beta)x_T$ is approximately equal to one'. We now first introduce the conditional distributions of the 2IP and the SPL estimator in the general case and then give the definition capturing what we just illustrated for the AR(1).\\
Let $m_T$ be a sequence of normalizing constants with $m_T \to \infty$ (e.g.\ $m_T=\sqrt{T}$). For any $t_1 \geq 1$, we define the sub $\sigma$-algebra $\mathcal{I}_{t_1:T} = \sigma(X_t: t_1 \leq t \leq T)$. We denote the \textit{conditional cdfs} of the 2IP and SPL estimator by
\begin{align}
\label{eq:2.4.2}
F_T^{2IP}(\tau|\mathcal{I}_{1:T}):= &\PP\big[m_T \big(\hat{\psi}_{T+1}(\X_{1:T},\Y_{1:T})-\psi_{T+1}\big) \leq \tau \:\big|\mathcal{I}_{1:T} \big]\\
\label{eq:2.4.3}
F_T^{SPL}(\tau|\mathcal{I}_{T_P:T}):= &\PP\big[m_T\big(\hat{\psi}_{T+1}(\X_{T_P:T},\X_{1:T_E}) - \psi_{T+1}\big) \leq \tau\:\big|\mathcal{I}_{T_P:T}\big]\:,
\end{align}
respectively, so that by specifying an event of $\mathcal{I}_{1:T}$ and $\mathcal{I}_{T_P:T}$, we see that (\ref{eq:2.4.2}) and (\ref{eq:2.4.3}) are just the centered and scaled distributions of (\ref{eq:2.3.11}) and (\ref{eq:2.3.16}), respectively. Please note that (\ref{eq:2.4.3}) actually also depends on $c$, see (\ref{eq:2.3.16}), but since our assumptions will ensure that this dependence vanishes asymptotically we prefer to suppress the dependence on $c$ here.

\begin{remark} Although not explicitly mentioned above we consider (\ref{eq:2.4.2}) and (\ref{eq:2.4.3}) to be regular conditional cdfs, which indicates that we assume that $F_T^{2IP}( \cdot|\mathcal{I}_{1:T})(\omega)$ and  $F_T^{SPL}( \cdot | \mathcal{I}_{T_p:T})(\omega)$ are cdfs for every $\omega \in \Omega$; for the exact definition and the existence see \citeauthor{dudley2002real} (\citeyear{dudley2002real}, Section 10.2).
\end{remark}
We can now define merging in probability (we do so without explicitly using the conditional cdfs of the 2IP and the SPL approach).
 \begin{definition}\textit{(Merging in Probability)} \label{def:2.2}
Two sequences of conditional cdfs $\{F_T\}$ and $\{G_T\}$ are said to \textit{merge in probability} if and only if $d_{BL}(F_T,G_T) \overset{p}{\to} 0$ as $T\to \infty$, where ``$\overset{p}{\to}$" denotes ``convergence in probability".
\end{definition}

\subsection{Merging of 2IP and SPL in Probability}\label{sec:2.4.2}
Here, we give conditions such that the conditional cdfs of the  2IP and SPL estimator merge. Clearly, the conditional confidence intervals are functions of these distributions so that their merging is a building block for the study of the conditional confidence intervals based on them. The conditions we give are divided into three parts. Roughly speaking, the first part (general assumptions) makes sure that the function we want to predict is well behaved and that we can estimate the parameter it depends on. The second part (two independent processes) and third part (SPL estimator) guarantee that these assumptions are met by the 2IP and the SPL method. To write the conditions in compact form we employ the usual stochastic order symbols $O_p$ and $o_p$. We assume that $\theta_0$ belongs to the interior of $\Theta$, i.e.\ $\theta_0 \in \mathring{\Theta}$, and we denote the set of all bounded, real-valued Lipschitz functions on $\R^r$ by $BL=\big \{h:\R^r\to \R:||h||_{BL}<\infty\big \}$. We start with the general assumptions.

\begin{assumption} (\textit{General Assumptions})
\label{as:2.0}
\begin{enumerate}[\ref{as:2.0}.a]
\item (\textit{Estimator})  $m_T \big(\hat{\theta}(\X_{1:T}) - \theta_0\big)\overset{d}{\to} G_{\infty}$  as $T \to \infty$ for some cdf $G_{\infty}: \mathbb{R}^r \rightarrow [0,1]$;

\item (\textit{Differentiability}) $\psi(\:\cdot\:; \theta)$ is continuous on $\Theta$ and twice differentiable on $\mathring{\Theta}$;

\item (\textit{Gradient}) $\Big|\Big|\frac{\partial \psi (X_T, X_{T-1}, \ldots ;\theta_0)}{\partial \theta}\Big|\Big|=O_{p}(1)$;

\item (\textit{Hessian}) $\sup_{\theta \in \mathscr{V}(\theta_0)}\Big|\Big|\frac{\partial^2 \psi (X_T, X_{T-1}, \ldots ;\theta)}{\partial \theta \partial \theta'}\Big|\Big|=O_{p}(1)$ for some open neighborhood $\mathscr{V}(\theta_0)$ around $\theta_0$;

\item (\textit{Initial Condition}) Given sequences $\{s_t\}$ and $\{c_t\}$, we have
\begin{align*}
m_T\big(\psi_{T+1}^s (\X_{t_1:T}^c; \theta_0) - \psi (X_T, X_{T-1}, \ldots ;\theta_0)\big)=o_{p}(1),\\
\bigg|\bigg|\frac{\partial \psi_{T+1}^s (\X_{t_1:T}^c; \theta_0)}{\partial \theta} - \frac{\partial \psi (X_T, X_{T-1}, \ldots ;\theta_0)}{\partial \theta}\bigg|\bigg|=o_{p}(1),\\
\sup_{\theta \in \mathscr{V}(\theta_0)}\bigg|\bigg|\frac{\partial^2 \psi_{T+1}^s (\X_{t_1:T}^c; \theta)}{\partial \theta \partial \theta'} - \frac{\partial^2 \psi (X_T, X_{T-1}, \ldots ;\theta)}{\partial \theta \partial \theta'}\bigg|\bigg|=o_{p}(1)
\end{align*}
for any $t_1 \geq 1$ such that $(T-t_1) / l_T \rightarrow \infty$ as $T \to \infty$ and for some model-specific $l_T$ with $l_T \rightarrow \infty$.
\end{enumerate}
\end{assumption}

Assumption \ref{as:2.0}.a implies the existence of a limiting distribution for the parameter estimator. The differentiability assumption in \ref{as:2.0}.b plus the boundedness Assumptions \ref{as:2.0}.c ensure that the scaled prediction function estimators can accurately be approximated by a Taylor expansion; see Lemma \ref{lem:2.1} for details. Assumption \ref{as:2.0}.e with $t_1 = 1$ ensures the negligibility of the starting values when using the full-sample for prediction, while taking $t_1 = T_P$ ensures that this extends to the case where additionally $X_1,\ldots,X_{T_P-1}$ are replaced by constants, i.e.~where only the subsample $(X_{T_P},\ldots,X_{T})$ is used for prediction. This assumption implicitly limits the choice of $T_P$; as replacing past values of $X_t$ for $t < T_P$ by arbitrary constants should have a negligible effect, $T - T_P$ needs to increase faster than some lower bound $l_T$. For models exhibiting an exponential decay in memory, it typically suffices to take $l_T = \log T$ (see e.g.~ \citeauthor{beutner2019technical}, \citeyear{beutner2019technical}, eq.~(4.6)).\\
For the 2IP estimator, we additionally need the two-independent-processes assumption, which is formalized in Assumption \ref{as:2.1}.
\begin{assumption} (\textit{Two Independent Processes}) \label{as:2.1}
\quad
\begin{enumerate}[\ref{as:2.1}.a]
\item (\textit{Existence}) $\{Y_t\}$ is a process defined on $(\Omega,\mathcal{F},\PP)$, distributed as $\{X_t\}$;

\item(\textit{Independence}) $\{Y_t\}$ is independent of $\{X_t\}$.
\end{enumerate}
\end{assumption}

For the SPL estimator we replace the two-independent-processes assumption by a stationarity and a weak dependence condition, which allows to split our sample into two (asymptotically) independent and identical subsamples. In addition we need an assumption on $T_P$ and $T_E$ as functions of $T$, that is $T_E (T)$ and $T_P (T)$.

\begin{assumption} \textit{(SPL Estimator)}
\label{as:2.2}
\begin{enumerate}[\ref{as:2.2}.a]
\item (\textit{Rates}) The functions $T_P:\mathbb{N} \to \mathbb{N}$ and $T_E:\mathbb{N} \to \mathbb{N}$ satisfy $T_E (T) < T_P(T)$ for all $T$, while $\frac{T - T_P (T)}{l_T} \rightarrow \infty$ and $m_{T_E (T)}/ m_T \rightarrow 1$ as $T \rightarrow \infty$;
\item (\textit{Strict Stationarity}) $\{X_t\}$ is a strictly stationary process;
\item (\textit{Weak Dependence}) $\{X_t\}$ satisfies for each $h \in BL$
\begin{align*}
\int h \:d \Big(G_{T_E}^{SPL}(\cdot|\mathcal{I}_{T_P:T})- G_{T_E}^{SPL}\Big) \overset{p}{\to}0 \qquad \text{as}\qquad T \to \infty,
\end{align*}
where $G_{T_E}^{SPL}$ denotes the unconditional cdf of $m_{T_E} \big(\hat{\theta}(\X_{1:T_E}) - \theta_0\big)$ and $G_{T_E}^{SPL}(\cdot|\mathcal{I}_{T_P:T})$ the corresponding conditional cdf given $\mathcal{I}_{T_P:T}$.
\end{enumerate}
\end{assumption}

The subsample size assumption in \ref{as:2.2}.a ensures that the number of observations used for conditioning is increasing, which along with the negligibility of the initial conditions implies that the truncation of the prediction function is negligible. Furthermore, the sample size used for estimation should increase fast enough that the respective scaling of the 2IP and SPL estimators, $m_{T}$ and $m_{T_E}$ respectively, are asymptotically identical. If $m_T$ increases no faster than a polynomial rate, which is generally the case, it is sufficient that $T_E/T \rightarrow 1$ for $m_{T_E}/m_T \rightarrow 1$ to hold. 

The stationarity assumption in \ref{as:2.2}.b can actually be relaxed; what matters is that the conditions in Assumption \ref{as:2.0} are still true if only a subsample is considered. In particular, we need that $m_{T_E} \big(\hat{\theta}(\X_{1:T_E}) - \theta_0\big)\overset{d}{\to} G_{\infty}$, which - along with the assumptions on gradient and Hessian - is certainly satisfied under stationarity. However, in general the assumption will be far too strict; here we use it simply to have a clear, interpretable assumption rather than a list of high-level assumptions that are difficult to interpret.
The weak dependence condition in \ref{as:2.2}.c is met by numerous Markov processes. Intuitively, $(X_1,\dots,X_{T_E})$ and $(X_{T_P},\dots,X_T)$ approach independence as their temporal distance $T_P-T_E$ increases. We illustrate a particular case in the Remark \ref{rem:2.2}.

\begin{remark}
\label{rem:2.2}
Suppose $\{X_t\}$ is strong mixing (cf.\ \citeauthor{doukhan1994mixing}, \citeyear{doukhan1994mixing}) and let $\alpha$ denote the strong mixing coefficient. For $h \in BL$ and for all $\epsilon>0$, Markov's and Ibragimov's inequality (cf.\ \citeauthor{hall2014martingale}, \citeyear{hall2014martingale}, Theorem A.5) imply
\begin{align*}
&\PP\bigg[\bigg|\int h\:d\Big( G_{T_E}^{SPL}(\cdot|\mathcal{I}_{T_P:T})- G_{T_E}^{SPL}\Big)\bigg|\geq \epsilon\bigg]\leq \frac{1}{\epsilon}\EE\bigg|\int h\:d\Big( G_{T_E}^{SPL}(\cdot|\mathcal{I}_{T_P:T})- G_{T_E}^{SPL}\Big)\bigg|\\
&=\frac{1}{\epsilon} \Cov\bigg[h\Big(m_{T_E} \big(\hat{\theta}(\X_{1:T_E}) - \theta_0\big)\Big),\text{sign}\Big\{\int h\:d\Big( G_{T_E}^{SPL}(\cdot|\mathcal{I}_{T_P:T})- G_{T_E}^{SPL}\Big)\Big \}\bigg]\\
&\leq  \frac{4||h||_{BL}}{\epsilon}\alpha(T_P - T_E)\:.
\end{align*}
Taking $T_P - T_E \to \infty$ such that $\alpha(T_P - T_E)\to 0$ verifies Assumption \ref{as:2.2}.c.
\end{remark}

Assumptions \ref{as:2.0} to \ref{as:2.2} are met by the AR and GARCH processes considered in Examples \ref{ex:2.1} and \ref{ex:2.2} (with 1.e holding for bounded sequences). A detailed verification of each assumption under mild conditions is provided in \cite{beutner2019technical}. We state the following theorem.
\begin{theorem}{(Merging of 2IP and SPL)}
\label{thm:2.1}
Under Assumptions \ref{as:2.0} to \ref{as:2.2},  $F_T^{2IP}(\cdot|\mathcal{I}_{1:T})$ and $F_T^{SPL}(\cdot|\mathcal{I}_{T_P:T})$ merge in probability.
\end{theorem}
Having established asymptotic closeness between the conditional cdfs $F_T^{2IP}(\cdot|\mathcal{I}_{1:T})$ and $F_T^{SPL}(\cdot|\mathcal{I}_{T_P:T})$, we now turn to the construction of asymptotic intervals.

\subsection{Interval Construction}\label{sec:2.4.3}

Henceforth, for any cdf $F$ we write $F^{-1}$ to denote its generalized inverse given by $F^{-1}(u)=\inf\big\{\tau \in \R:F(\tau)\geq u\big\}$. A confidence interval for $\psi_{T+1}$ based on quantiles of \eqref{eq:2.4.2} or \eqref{eq:2.4.3} is typically infeasible as these cumulative distribution functions are unknown for finite $T$. Here, they are infeasible because roughly they are the distribution functions of some weights which induce merging multiplied by $m_T \big(\hat{\theta}(\X_{1:T}) - \theta_0\big)$ and $m_{T_E} \big(\hat{\theta}(\X_{1:T_E}) - \theta_0\big)$, respectively, where, in general, the distributions of $m_T \big(\hat{\theta}(\X_{1:T}) - \theta_0\big)$ and $m_{T_E} \big(\hat{\theta}(\X_{1:T_E}) - \theta_0\big)$ are unknown in finite samples. Since these are the only unknown distributions an asymptotic approximation can be based on $G_{\infty}$ with merging induced by the non-convergent weights. In general, we also need to estimate $G_{\infty}$; see Examples \ref{example estimating G infty by parameters} and \ref{example approx G infty bootstrap} below for common approaches. We denote estimators of \eqref{eq:2.4.2} and \eqref{eq:2.4.3} resulting from this approximation by $\widehat{F_T^{2IP}}(\cdot)$ and $\widehat{F_T^{SPL}}(\cdot)$, respectively. In Section \ref{sec:2.4.4}, we provide explicit expressions when $G_{\infty}$ is multivariate normal. For the general construction, we refer to relations \eqref{eq:2.A.3} and \eqref{eq:2.A.4} in Appendix \ref{app:A} and the explanations preceding these relations. Based on the 2IP approach, we consider an interval of the form
\begin{align}
\label{eq:2.4.8}
 I_{\gamma}^{2IP}(\x_{1:T},\Y_{1:T}) =&\bigg[\hat{\psi}_{T+1}(\x_{1:T},\Y_{1:T})-\frac{\widehat{F_T^{2IP}}^{-1}(1-\gamma_2)}{m_T}\:,\:\hat{\psi}_{T+1}(\x_{1:T},\Y_{1:T})-\frac{\widehat{F_T^{2IP}}^{-1}(\gamma_1)}{m_T}\bigg] \nonumber \\
\stackrel{(\ref{eq:2.3.11})}{=}&\bigg[\hat{\psi}_{T+1|1:T}^{2IP}-\frac{\widehat{F_T^{2IP}}^{-1}(1-\gamma_2)}{m_T}\:,\:\hat{\psi}_{T+1|1:T}^{2IP}-\frac{\widehat{F_T^{2IP}}^{-1}(\gamma_1)}{m_T}\bigg]\:,
\end{align}
where $\gamma_1,\gamma_2 \in [0,1)$ satisfy $\gamma=\gamma_1+\gamma_2$. We typically take $\gamma_1=\gamma_2=\gamma/2$ such that the interval is equal-tailed. Similarly, we construct the following sample split interval:
\begin{align}
\label{eq:2.4.9}
\begin{split}
I_{\gamma}^{SPL}(\x_{T_P:T}^c,\X_{1:T_E})
=&\bigg[\hat{\psi}_{T+1}(\x_{T_P:T}^c,\X_{1:T_E})-\frac{\widehat{F_T^{SPL}}^{-1}(1-\gamma_2)}{m_T}\:,\:\hat{\psi}_{T+1}(\x_{T_P:T}^c,\X_{1:T_E})-\frac{\widehat{F_T^{SPL}}^{-1}(\gamma_1)}{m_T}\bigg]\\
\stackrel{(\ref{eq:2.3.16})}{=}&\bigg[\hat{\psi}_{T+1|T_P:T}^{SPL}-\frac{\widehat{F_T^{SPL}}^{-1}(1-\gamma_2)}{m_T}\:,\:\hat{\psi}_{T+1|T_P:T}^{SPL}-\frac{\widehat{F_T^{SPL}}^{-1}(\gamma_1)}{m_T}\bigg]\:.
\end{split}
\end{align}
To achieve correct coverage, we need that $\widehat{F_T^{2IP}}(\cdot)$ and $F_T^{2IP}(\cdot|\mathcal{I}_{1:T})$ merge in probability and likewise for SPL. A sufficient condition for this in our setting is that we can consistently estimate the asymptotic distribution of the parameter estimator, $G_{\infty}$, by an appropriate estimator. This is formulated in Assumption \ref{as:2.4} below.
\begin{assumption}{\textit{(CDF Estimator)}} \label{as:2.4} Let $\widehat{G}_T (\cdot)$ denote a random ($r$-dimensional) cdf as a function of $\X_{1:T}$, used to estimate $G_{\infty}$. Then $\int h \:d \widehat{G}_T (\cdot) \overset{p}{\to}\int h \:d G_{\infty}$ as $T \to \infty$ for all $h \in BL$.
\end{assumption}
\noindent
Although we did not explicitly specify in Assumption \ref{as:2.4} the dependence of $\widehat{G}_T$ on $\X_{1:T}$, it should be understood to hold for any subsample of $\X_{1:T}$ whose size goes to infinity. The verification of Assumption \ref{as:2.4} is a standard step in asymptotic analysis. The two examples below provide common methods for verifying Assumption \ref{as:2.4}.
\begin{example}\label{example estimating G infty by parameters}
Suppose that $G_{\infty}$ belongs to some parametric family $\{G_{\theta,\xi} | \theta \in \Theta, \xi \in \Xi\}$. Then, given some consistent estimators $\hat{\theta}(\X_{1:T})$ and $\hat{\xi}(\X_{1:T})$ for $\theta_0$ and $\xi_0$ respectively, it follows from the continuous mapping theorem that $\widehat{G}_T = G_{\hat{\theta}(\X_{1:T}), \hat{\xi}(\X_{1:T})}$ satisfies Assumption \ref{as:2.4} if $G_{\theta,\xi}$ is continuous in $\theta$ and $\xi$.
\end{example}
\begin{example}\label{example approx G infty bootstrap}
If $\hat{G}_T$ is based on a consistent bootstrap procedure for $G_{\infty}$ then Assumption \ref{as:2.4} clearly holds. 
\end{example}

The following theorem states the intervals' asymptotic validity.
\begin{theorem}{ (Asymptotic Coverage)}
\label{thm:2.2}
\begin{enumerate}
\item 	\begin{enumerate}[(a)]
              \item  Under Assumption \ref{as:2.0}, \ref{as:2.1} and \ref{as:2.4}, $F_T^{2IP}(\cdot|\mathcal{I}_{1:T})$ and $\widehat{F_T^{2IP}}(\cdot)$ merge in probability.
              \item
     If in addition $\widehat{F_T^{2IP}}(\cdot)$ is stochastically uniformly equicontinuous, then
  \begin{align}
  \label{eq:2.4.10}
  \PP \Big[  I_\gamma^{2IP}(\x_{1:T},\Y_{1:T}) \ni \psi_{T+1}  \Big|\mathcal{I}_{1:T}\Big]\overset{p}{\to} 1 -\gamma\:.
  \end{align}
      \end{enumerate}
\item \begin{enumerate}[(a)]
        \item Under Assumption \ref{as:2.0}, \ref{as:2.2} and \ref{as:2.4}, $F_T^{SPL}(\cdot|\mathcal{I}_{T_P:T})$ and $\widehat{F_T^{SPL}}(\cdot)$ merge in probability.

        \item If in addition $\widehat{F_T^{SPL}}(\cdot)$ is stochastically uniformly equicontinuous, then
    \begin{align}
    \label{eq:2.4.11}
    \PP \Big[ I_\gamma^{SPL}(\x_{T_P:T}^c,\X_{1:T_E}) \ni \psi_{T+1} \Big|\mathcal{I}_{T_P:T}\Big]\overset{p}{\to} 1 -\gamma\:.
    \end{align}
    \end{enumerate}
\end{enumerate}
\end{theorem}
However, the standard approach, motivated by $I_\gamma^{2IP}$ as in \eqref{eq:2.4.10}, computes an interval of the form $I_\gamma^{STA}(\x_{1:T},\x_{1:T}) = I_\gamma^{2IP}(\x_{1:T},\x_{1:T})$ as only one sample realization is available. This, of course, strongly violates the independence assumption of $\{X_t\}$ and $\{Y_t\}$. Specifically, replacing $\Y_{1:T}$ by $\X_{1:T}$ in equation \eqref{eq:2.4.8}, leads to
\begin{align}
\label{eq:2.4.12}
I_{\gamma}^{STA*}(\x_{1:T},\X_{1:T})
=&\bigg[\hat{\psi}_{T+1}(\x_{1:T},\X_{1:T})-\frac{\widehat{F_T^{STA}}^{-1}(1-\gamma_2)}{m_T}\:,\:\hat{\psi}_{T+1}(\x_{1:T},\X_{1:T})-\frac{\widehat{F_T^{STA}}^{-1}(\gamma_1)}{m_T}\bigg] \nonumber \\
\stackrel{(\ref{eq:2.3.7})}{=}&\bigg[\hat{\psi}_{T+1|1:T}^{STA*}-\frac{\widehat{F_T^{STA}}^{-1}(1-\gamma_2)}{m_T}\:,\:\hat{\psi}_{T+1|1:T}^{STA*}-\frac{\widehat{F_T^{STA}}^{-1}(\gamma_1)}{m_T}\bigg]\:,
\end{align}
where $\widehat{F_T^{STA}}(\cdot)$ is defined in relation \eqref{eq:2.A.5} and the text preceding it. Whereas it is difficult to justify a conditional confidence interval like $I_{\gamma}^{STA}(\x_{1:T},\X_{1:T})$ directly due to the lack of randomness, we can provide a justification by characterizing how closely the interval resembles the SPL interval. We establish the asymptotic equivalence, defined in terms of location and (scaled) length, between the two intervals in the following theorem. Note that, as our characterization of equivalence is probabilistic, we need to introduce the ``doubly random'' versions of the STA and SPL estimators, where the sample we condition on is considered random. These estimators are denoted as $\hat\psi_{T+1} (\X_{1:T}, \X_{1:T})$ and $\hat\psi_{T+1} (\X_{T_P:T}^c, \X_{1:T_E})$ respectively.
\begin{theorem} (Asymptotic Equivalence Confidence Intervals)
\label{thm:2.3}
\begin{enumerate}
	\item (Location) If Assumptions 1-3 hold, then $\hat\psi_{T+1} (\X_{1:T}, \X_{1:T}) - \hat\psi_{T+1} (\X_{T_P:T}^c, \X_{1:T_E}) \overset{p}{\to} 0$.
    \item (Length) Under the assumptions of Theorem \ref{thm:2.1} and \ref{thm:2.2} and  $\widehat{F_T^{SPL}}^{-1}(\cdot)$ being stochastically pointwise continuous at $u =\gamma_1,1-\gamma_2$, we have
\begin{align}
\label{eq:2.4.13}
    \widehat{F_T^{STA}}^{-1}(u)-\widehat{F_T^{SPL}}^{-1}(u)\overset{p}{\to}0\:.
\end{align}
\end{enumerate}
\end{theorem}

The first implication states that the locations of the two intervals coincide asymptotically. The second statement establishes asymptotic closeness of the selected quantiles such that the scaled lengths of the intervals in \eqref{eq:2.4.9} and \eqref{eq:2.4.12} coincide asymptotically. As such, our sample-split interval coincides asymptotically with the standard interval, meaning that the standard interval can be substituted for an (asymptotically) equivalent interval which has a formal justification in terms of conditional coverage. As such, this provides a justification for the intervals commonly constructed in practice without having to rely on the two-independent-processes assumption.

\subsection{Interval Construction Under Normality}
\label{sec:2.4.4}

In this subsection we present intervals of reduced form and simplified theoretical results under asymptotic normality of the parameter estimator.
\begin{assumption}{\textit{(Normality)}}
\label{as:2.5}
Let $G_{\infty}$ be the cdf of the $N(0,\Upsilon_0)$ distribution with $\Upsilon_0=\Upsilon(\theta_0,\xi_0)$ and assume there exist $\hat{\Upsilon}(\X_{1:T})$ converging in probability to $\Upsilon_0$.
\end{assumption}
\noindent
Usually, the covariance estimator is obtained by inserting consistent estimators for $\theta_0$ and $\xi_0$ into $\Upsilon_0$. Following the plug-in principle, we estimate $F_T^{2IP}(\cdot|\mathcal{I}_{1:T})$ by a normal distribution with mean $0$ and variance $\hat{\upsilon}_T^{2IP}=\frac{\partial \psi_{T+1}(\x_{1:T};\hat{\theta}(\Y_{1:T}))}{\partial \theta'}\hat{\Upsilon}(\Y_{1:T})\frac{\partial \psi_{T+1}(x_{1:T};\hat{\theta}(\Y_{1:T}))}{\partial \theta}$ such that $\widehat{F_T^{2IP}}(\cdot)=\Phi\big(\cdot/\sqrt{\hat{\upsilon}_T^{2IP}}\big)$. Then, the interval in \eqref{eq:2.4.8} simplifies to
\begin{align}
\label{eq:2.4.14}
I_{\gamma}^{2IP}(\x_{1:T},\Y_{1:T}) =&\Bigg[\hat{\psi}_{T+1|1:T}^{2IP}-\frac{\sqrt{\hat{\upsilon}_T^{2IP}}\Phi^{-1}(1-\gamma_2)}{m_T}\:,\:\hat{\psi}_{T+1|1:T}^{2IP}-\frac{\sqrt{\hat{\upsilon}_T^{2IP}}\Phi^{-1}(\gamma_1)}{m_T}\Bigg]\:.
\end{align}
Similarly, for the sample-split approach we consider  $\widehat{F_T^{SPL}}(\cdot)= \Phi\big(\cdot / \sqrt{\hat{\upsilon}_T^{SPL}}\big)$ with $\hat{\upsilon}_T^{SPL}=\frac{\partial \psi_{T+1}(\x_{T_P:T}^{c};\hat{\theta}(\X_{1:T_E}))}{\partial \theta'}\hat{\Upsilon}(\X_{1:T_E})\frac{\partial \psi_{T+1}(\x_{T_P:T}^{c};\hat{\theta}(\X_{1:T_E}))}{\partial \theta'}$ such that \eqref{eq:2.4.9} reduces to
\begin{align}
\label{eq:2.4.15}
\begin{split}
I_{\gamma}^{SPL}(\x_{T_P:T},\X_{1:T_E})
=\bigg[\hat{\psi}_{T+1|T_P:T}^{SPL}-\frac{\sqrt{\hat{\upsilon}_T^{SPL}}\Phi^{-1}(1-\gamma_2)}{m_T}\:,\:\hat{\psi}_{T+1|T_P:T}^{SPL}-\frac{\sqrt{\hat{\upsilon}_T^{SPL}}\Phi^{-1}(\gamma_1)}{m_T}\bigg]\:.
\end{split}
\end{align}
In Appendix B we show that if the variance estimator is bounded away from zero in probability, e.g. $1/\hat{\upsilon}_T^{2IP}=O_p(1)$, then $\widehat{F_T^{2IP}}(\cdot)$ is stochastically uniform equicontinuous. Therefore, the asymptotic validity of both intervals can be deduced from Theorem \ref{thm:2.2}.
\begin{corollary}{(Asymptotic Coverage under Normality)}
\label{cor:2.1}
\begin{enumerate}
\item 	\begin{enumerate}[(a)]
              \item  Under Assumption \ref{as:2.0}, \ref{as:2.1} and \ref{as:2.5}, $F_T^{2IP}(\cdot|\mathcal{I}_{1:T})$ and $\Phi\big(\cdot/ \sqrt{\hat{\upsilon}_T^{2IP}}\big)$ merge in probability.
              \item
     If in addition $1/\hat{\upsilon}_T^{2IP}=O_p(1)$, $\PP \Big[  I_\gamma^{2IP}(\x_{1:T},\Y_{1:T}) \ni \psi_{T+1}   \Big|\mathcal{I}_{1:T}\Big]\overset{p}{\to} 1 -\gamma$.
      \end{enumerate}
\item \begin{enumerate}[(a)]
        \item Under Assumption \ref{as:2.0}, \ref{as:2.2} and \ref{as:2.5}, $F_T^{SPL}(\cdot|\mathcal{I}_{T_P:T})$ and $\Phi\big(\cdot/\sqrt{\hat{\upsilon}_T^{SPL}}\big)$ merge in probability.

        \item If in addition $1/\hat{\upsilon}_T^{SPL}=O_p(1)$, $\PP \Big[ I_\gamma^{SPL}(\x_{T_P:T}^{c},\X_{1:T_E}) \ni  \psi_{T+1} \Big|\mathcal{I}_{T_P:T}\Big]\overset{p}{\to} 1 -\gamma$.
    \end{enumerate}
\end{enumerate}
\end{corollary}
\noindent
Bounding the variance estimator away from zero in probability to establish that the conditional coverage probability converges to $1-\gamma$ in probability has intuitive appeal: as $\hat{\upsilon}_T^{2IP}$ approaches zero,  $N\big(0,\hat{\upsilon}_T^{2IP}\big)$ becomes degenerate while the interval in \eqref{eq:2.4.14} collapses (similar for SPL).

For the standard interval, replacing $\Y_{1:T}$ by $\X_{1:T}$ in \eqref{eq:2.4.14} leads to
\begin{align}
\label{eq:2.4.18}
I_{\gamma}^{STA *}(\x_{1:T},\X_{1:T}) =&\bigg[\hat{\psi}_{T+1|T_P:T}^{STA*}-\frac{\sqrt{\hat{\upsilon}_{T}^{STA}}\Phi^{-1}(1-\gamma_2)}{m_T}\:,\:\hat{\psi}_{T+1|T_P:T}^{STA*}-\frac{\sqrt{\hat{\upsilon}_{T}^{STA}}\Phi^{-1}(\gamma_1)}{m_T}\bigg]
\end{align}
with $\hat{\upsilon}_{T}^{STA}=\frac{\partial \psi_{T+1}(x_{1:T};\hat{\theta}(\X_{1:T}))}{\partial \theta'}\hat{\Upsilon}(\mathbf{X}_n)\frac{\partial \psi_{T+1}(\x_{1:T};\hat{\theta}(\X_{1:T}))}{\partial \theta}$.
In Appendix B we prove that $\hat{\upsilon}_T^{SPL}$ is bounded in probability, which in turn implies that the quantile function $\widehat{F_T^{SPL}}^{-1}(\cdot)=\sqrt{\hat{\upsilon}_T^{SPL}}\Phi^{-1}(\cdot)$ is stochastically pointwise equicontinuous at any $u \in \R$. Hence, Theorem \ref{thm:2.3} applies. Whereas the first statement of the theorem remains unaffected, its second statement reads as follows under normality.
\begin{corollary}{(Length under Normality)}
\label{cor:2.2}
Under the assumptions of Theorem \ref{thm:2.1} and Corollary \ref{cor:2.1}, we have $\sqrt{\hat{\upsilon}_{T}^{STA}}\Phi^{-1}(u)-\sqrt{\hat{\upsilon}_T^{SPL}}\Phi^{-1}(u)\overset{p}{\to}0$
for $u =\gamma_1,1-\gamma_2$.
\end{corollary}

\section{Prediction Intervals}
\label{sec:2.5}

The preceding sections have focused purely on the construction of conditional confidence intervals to account for parameter uncertainty. Regarding prediction, a second source of uncertainty arises, that corresponds to the model's innovation process. In this setting, parameter estimation is typically disregarded in textbooks as the stochastic fluctuation stemming from the estimation procedure is generally dominated by the stochastic fluctuation of the innovations. Although the resulting prediction intervals may be asymptotically valid, they are typically characterized by under-coverage in finite samples. In response, \cite{pan2016abootstrap} introduce the concept of \textit{asymptotic pertinence} to evaluate distribution approximations that account for the two sources of randomness, innovation and parameter estimation uncertainty, according to their general orders of magnitude.
Whereas \cite{kunitomo1985properties} and \cite{samaranayake1988properties} study properties of the \textit{unconditional} law of the forecast error, we focus on its \textit{conditional} distribution to conform with the relevance property of \cite{kabaila1999relevance}. The fundamental issue also arises when considering prediction if one attempts to account for parameter uncertainty. To illustrate this point, we revisit the introductory examples and write $*$ to denote the convolution operator.\footnote{For independent variables $X$ and $Y$ with $X\sim F_X$, $Y\sim F_Y$ and $Z=X+Y \sim F_{Z}$, we write $F_{Z}=F_X * F_Y$ to denote $F_{Z}(z) = \int_{-\infty}^zF_X(z-y)dF_Y(y)$.}
\begin{Example} \textbf{\ref{ex:2.1}.} \textit{(continued)}
Prediction intervals for the AR are often constructed around the point estimate for the conditional mean. The conditional distribution of the forecast error decomposes into
\begin{align}
\label{eq:2.5.1}
\begin{split}
\PP\big[X_{T+1}-\hat{\beta}(\mathbf{X}_{1:T})X_T\leq \cdot|X_T=x_T\big]
=&\PP\big[\beta X_T-\hat{\beta}(\mathbf{X}_{1:T})X_T\leq \cdot|X_T=x_T\big]\\
&*\PP\big[\varepsilon_{T+1}\leq \cdot\big]\:,
\end{split}
\end{align}
corresponding to estimation and innovation uncertainty, respectively. As argued above, an approximation of 
$ \PP\big[\beta X_T-\hat{\beta}(\mathbf{X}_{1:T})X_T\leq \cdot|X_T=x_T\big]$ appears rather cumbersome. In the special case of $\varepsilon_t\overset{iid}{\sim}N\big(0,\sigma^2_\varepsilon\big)$,  \citeauthor{phillips1979sampling} (\citeyear{phillips1979sampling}, Thm.\ 3) derives an approximation for \eqref{eq:2.5.1} based on Edgeworth expansions.
\end{Example}
\begin{Example} \textbf{\ref{ex:2.2}.} \textit{(continued)}
Suppose we are interested in providing a prediction interval for $X_{T+1}^2$ in the GARCH(1,1). Conditioning on $\mathbf{X}_{1:T}=\mathbf{x}_{1:T}$, a natural estimate of $X_{T+1}^2$ is $\hat{\sigma}_{T+1|1:T}^{2\;STA}$ as defined in \eqref{eq:2.2.9}, since $\sigma_{T+1|1:T}^2$ is its expected value given information up to time $T$. As
\begin{align}
\label{eq:2.5.3}
\begin{split}
\PP\big[X_{T+1}^2-\hat{\sigma}_{T+1}^{2\;STA*}\leq \cdot|\mathbf{X}_{1:T}=\mathbf{x}_{1:T}\big]
=&\PP\big[\sigma_{T+1}^2-\hat{\sigma}_{T+1}^{2\;STA*}\leq \cdot|\mathbf{X}_{1:T}=\mathbf{x}_{1:T}\big]\\ &*\PP\big[\sigma_{T+1|1:T}^2(\varepsilon_{T+1}^2-1)\leq \cdot\big]\:,
\end{split}
\end{align}
where $\hat{\sigma}_{T+1}^{2\;STA*}$ is defined in \eqref{eq:2.2.10}, the desired prediction interval, say $J_\gamma^{STA}$, leads to a sensible probabilistic statement due to variability in $\varepsilon_{T+1}^2$:
\begin{align}
\label{eq:2.5.4}
\PP\Big[ X_{T+1}^2 \in J_\gamma^{STA}(\mathbf{X}_{1:T},\mathbf{X}_{1:T}) \Big|\mathbf{X}_{1:T}=\mathbf{x}_{1:T}\Big] \underset{(\approx)}{=} 1 -\gamma\:.
\end{align}
However, it cannot incorporate parameter uncertainty either, since the conditional distribution $\PP\big[\sigma_{T+1}^2-\hat{\sigma}_{T+1}^{2\;STA*}\leq \cdot|\mathbf{X}_{1:T}=\mathbf{x}_{1:T}\big]=\PP\big[\sigma_{T+1|1:T}^2-\hat{\sigma}_{T+1|1:T}^{2\;STA}\leq \cdot\big]$  is degenerate.
\end{Example}
In his textbook \cite{pesaran2015time} resorts to a Bayesian-akin approach to avoid the fundamental issue in forecasting. He argues that $\theta$, although ``fixed at the estimation stage ... is viewed best as a random variable at the forecasting stage" (p.\ 389). Consequently, he assigns some posterior distribution to $\theta$ motivated by an uninformed prior. Treating $\theta$ not fixed but random, the fundamental issue does not arise, however combining a frequentist view with a Bayesian method does not seem to be coherent.

 \cite{barndorff1996prediction} require the existence of a transitive statistic $U=U(\mathbf{X}_{1:T})$ of fixed low dimension to establish conditional independence between the sample $\mathbf{X}_{1:T}$ and their considered future random variable given $U=u$. \citeauthor{vidoni2004improved} (\citeyear{vidoni2004improved}, \citeyear{vidoni2009improved}, \citeyear{vidoni2009simple}, \citeyear{vidoni2016improved}), \cite{kabaila1999relevance}, \cite{kabaila2004adjustment},
and \citeauthor{kabaila2008improved} (\citeyear{kabaila2008improved}, \citeyear{kabaila2010asymptotic}) extend their approach and derive improved prediction intervals. Although these methods absorb an additional $O(T^{-1})$ term in the associated conditional coverage probability, there are several drawbacks associated with them: the innovation distribution needs typically be specified (e.g.\ Gaussian),  the results apply only to a limited set of estimators (e.g.\ maximum likelihood) and  their framework can only incorporate finite autoregressive components (e.g.\ AR$(p)$).

Assuming two independent processes with the same stochastic structure, using one for prediction and one for the estimation of the parameters, alleviates the fundamental issue faced in the continued Examples \ref{ex:2.1} and \ref{ex:2.2}. As the conditional distributions of the 2IP and SPL estimators merge in probability by Theorem \ref{thm:2.1}, the 2IP assumption can be avoided by following a sample-split approach as described in Section \ref{sec:2.3.3}.

\section{Conclusion} \label{sec:2.6}

In the paper at hand, we study the construction of confidence intervals for conditional objects such as conditional means or conditional variances, focusing on  the conceptual issue that arises in the process of taking parameter uncertainty into account. It stems from the fact that on one hand one must condition on the sample as the past informs about the present and future, yet on the other hand one must allow the data up to now to be treated as random to account for estimation uncertainty. Assuming two independent processes with the same stochastic structure, where one is used for conditioning and one for the estimation of the parameters, bypasses this issue, but the assumption itself  can generally not be justified in applications. To avoid this assumption, we propose a solution based on a simple sample-split approach, that requires a much more realistic weak dependence condition instead. To acknowledge that the conditional quantities vary over time, we employ a merging concept generalizing the notion of weak convergence. The conditional distributions of the sample-split estimator and the estimator based on the two-independent-processes assumption are shown to merge in probability under mild conditions. The corresponding sample-split intervals are shown to coincide asymptotically with the intervals commonly constructed by practitioners, which provides a novel and theoretically satisfactory justification for commonly constructed confidence intervals for conditional objects, applicable to a wide class of time series models, including ARMA and GARCH-type models.

One limitation to our approach is that we restrict ourselves to univariate time series and objects of interests. At the expense of more involved notation this could be readily extended to multivariate time series and objects of interests. A second, and more restrictive, limitation is our weak dependence assumption needed to achieve asymptotic independence between the two subsamples, which for instance rules out application to integrated processes. Given the fundamental role of this assumption in our setup, it appears difficult to generalize this. However, this also casts further doubt on the two-independent-processes assumption as validation for confidence intervals constructed for such persistent processes. A case-by-case treatment, as for instance done by \citet{gospodinov2002median} for near unit root processes and \cite{samaranayake1988properties} for explosive processes, appears to be necessary in such cases, and standard confidence intervals should be treated with caution.

\bibliographystyle{chicago}

\begin{thebibliography}{}

\bibitem[\protect\citeauthoryear{Akaike}{Akaike}{1969}]{akaike1969fitting}
Akaike, H. (1969).
\newblock Fitting autoregressive models for prediction.
\newblock {\em Annals of the Institute of Statistical Mathematics\/}~{\em
  21\/}(1), 243--247.

\bibitem[\protect\citeauthoryear{Barndorff-Nielsen and Cox}{Barndorff-Nielsen
  and Cox}{1996}]{barndorff1996prediction}
Barndorff-Nielsen, O.~E. and D.~R. Cox (1996).
\newblock Prediction and asymptotics.
\newblock {\em Bernoulli\/}~{\em 2\/}(4), 319--340.

\bibitem[\protect\citeauthoryear{Belyaev and Sj\"{o}stedt-De~Luna}{Belyaev and
  Sj\"{o}stedt-De~Luna}{2000}]{belyaev2000weakly}
Belyaev, Y. and S.~Sj\"{o}stedt-De~Luna (2000).
\newblock Weakly approaching sequences of random distributions.
\newblock {\em Journal of Applied Probability\/}~{\em 37\/}(3), 807--822.

\bibitem[\protect\citeauthoryear{Beutner, Heinemann, and Smeekes}{Beutner
  et~al.}{2017a}]{beutner2017justification}
Beutner, E., A.~Heinemann, and S.~Smeekes (2017a).
\newblock A justification of conditional confidence intervals.
\newblock GSBE Research Memorandum RM/17/023, Maastricht University.

\bibitem[\protect\citeauthoryear{Beutner, Heinemann, and Smeekes}{Beutner
  et~al.}{2017b}]{beutner2017technical}
Beutner, E., A.~Heinemann, and S.~Smeekes (2017b).
\newblock Technical report on ``a justification of conditional confidence
  intervals''.
\newblock Technical report, Maastricht University,
  \url{http://researchers-sbe.unimaas.nl/stephansmeekes/research/}.

\bibitem[\protect\citeauthoryear{Beutner, Heinemann, and Smeekes}{Beutner
  et~al.}{2019}]{beutner2019technical}
Beutner, E., A.~Heinemann, and S.~Smeekes (2019).
\newblock A general framework for prediction in time series models.
\newblock Working paper, Maastricht University,
  \url{http://researchers-sbe.unimaas.nl/stephansmeekes/research/}.

\bibitem[\protect\citeauthoryear{Billingsley}{Billingsley}{1986}]{billingsley1986probability}
Billingsley, P. (1986).
\newblock {\em Probability and Measure}.
\newblock New York: John Wiley \& Sons.

\bibitem[\protect\citeauthoryear{Blasques, Koopman, {\L}asak, and
  Lucas}{Blasques et~al.}{2016}]{blasques2016sample}
Blasques, F., S.~J. Koopman, K.~{\L}asak, and A.~Lucas (2016).
\newblock In-sample confidence bands and out-of-sample forecast bands for
  time-varying parameters in observation-driven models.
\newblock {\em International Journal of Forecasting\/}~{\em 32\/}(3), 875--887.

\bibitem[\protect\citeauthoryear{Bollerslev}{Bollerslev}{1986}]{bollerslev1986generalized}
Bollerslev, T. (1986).
\newblock Generalized autoregressive conditional heteroskedasticity.
\newblock {\em Journal of Econometrics\/}~{\em 31\/}(3), 307--327.

\bibitem[\protect\citeauthoryear{Boussama, Fuchs, and Stelzer}{Boussama
  et~al.}{2011}]{boussama2011stationarity}
Boussama, F., F.~Fuchs, and R.~Stelzer (2011).
\newblock Stationarity and geometric ergodicity of \uppercase{BEKK}
  multivariate \uppercase{Garch} models.
\newblock {\em Stochastic Processes and Their Applications\/}~{\em 121\/}(10),
  2331--2360.

\bibitem[\protect\citeauthoryear{Castillo and Rousseau}{Castillo and
  Rousseau}{2015}]{CastilloAoS15}
Castillo, I. and J.~Rousseau (2015).
\newblock A \uppercase{B}ernstein–von \uppercase{M}ises theorem for smooth
  functionals in semiparametric models.
\newblock {\em Annals of Statistics\/}~{\em 43\/}(6), 2353--2383.

\bibitem[\protect\citeauthoryear{Cavaliere, Georgiev, and Taylor}{Cavaliere
  et~al.}{2013}]{cavaliere2013wild}
Cavaliere, G., I.~Georgiev, and A.~M.~R. Taylor (2013).
\newblock Wild bootstrap of the sample mean in the infinite variance case.
\newblock {\em Econometric Reviews\/}~{\em 32\/}(2), 204--219.

\bibitem[\protect\citeauthoryear{D'Aristotile, Diaconis, and
  Freedman}{D'Aristotile et~al.}{1988}]{d1988merging}
D'Aristotile, A., P.~Diaconis, and D.~Freedman (1988).
\newblock On merging of probabilities.
\newblock {\em Sankhy{\=a}: The Indian Journal of Statistics, Series A\/}~{\em
  50\/}(3), 363--380.

\bibitem[\protect\citeauthoryear{Davydov and Rotar}{Davydov and
  Rotar}{2009}]{davydov2009asymptotic}
Davydov, Y. and V.~Rotar (2009).
\newblock On asymptotic proximity of distributions.
\newblock {\em Journal of Theoretical Probability\/}~{\em 22\/}(1), 82--98.

\bibitem[\protect\citeauthoryear{Doukhan}{Doukhan}{1994}]{doukhan1994mixing}
Doukhan, P. (1994).
\newblock {\em Mixing: Properties and Examples}.
\newblock New York: Springer.

\bibitem[\protect\citeauthoryear{Dudley}{Dudley}{2002}]{dudley2002real}
Dudley, R.~M. (2002).
\newblock {\em Real Analysis and Probability}.
\newblock Cambridge: Cambridge University Press.

\bibitem[\protect\citeauthoryear{Dufour and Taamouti}{Dufour and
  Taamouti}{2010}]{dufour2010short}
Dufour, J.-M. and A.~Taamouti (2010).
\newblock Short and long run causality measures: theory and inference.
\newblock {\em Journal of Econometrics\/}~{\em 154\/}(1), 42--58.

\bibitem[\protect\citeauthoryear{Engle}{Engle}{1982}]{engle1982autoregressive}
Engle, R.~F. (1982).
\newblock Autoregressive conditional heteroscedasticity with estimates of the
  variance of \uppercase{U}nited \uppercase{K}ingdom inflation.
\newblock {\em Econometrica\/}~{\em 50\/}(4), 987--1007.

\bibitem[\protect\citeauthoryear{Francq and Zako{\"\i}an}{Francq and
  Zako{\"\i}an}{2015}]{francq2015risk}
Francq, C. and J.-M. Zako{\"\i}an (2015).
\newblock Risk-parameter estimation in volatility models.
\newblock {\em Journal of Econometrics\/}~{\em 184\/}(1), 158--173.

\bibitem[\protect\citeauthoryear{Gospodinov}{Gospodinov}{2002}]{gospodinov2002median}
Gospodinov, N. (2002).
\newblock Median unbiased forecasts for highly persistent autoregressive
  processes.
\newblock {\em Journal of Econometrics\/}~{\em 111\/}(1), 85--101.

\bibitem[\protect\citeauthoryear{Hall and Heyde}{Hall and
  Heyde}{1980}]{hall2014martingale}
Hall, P. and C.~C. Heyde (1980).
\newblock {\em Martingale Limit Theory and Its Application}.
\newblock London: Academic Press.

\bibitem[\protect\citeauthoryear{Hamilton}{Hamilton}{1994}]{hamilton1994time}
Hamilton, J.~D. (1994).
\newblock {\em Time Series Analysis}.
\newblock Princeton: Princeton University Press.

\bibitem[\protect\citeauthoryear{Hansen}{Hansen}{2006}]{hansen2006interval}
Hansen, B.~E. (2006).
\newblock Interval forecasts and parameter uncertainty.
\newblock {\em Journal of Econometrics\/}~{\em 135\/}(1), 377--398.

\bibitem[\protect\citeauthoryear{Huber}{Huber}{2009}]{huber2009robust}
Huber, P.~J. (2009).
\newblock {\em Robust Statistics\/} (2nd ed.).
\newblock New York: John Wiley \& Sons.

\bibitem[\protect\citeauthoryear{Ing and Wei}{Ing and Wei}{2003}]{ing2003same}
Ing, C.-K. and C.-Z. Wei (2003).
\newblock On same-realization prediction in an infinite-order autoregressive
  process.
\newblock {\em Journal of Multivariate Analysis\/}~{\em 85\/}(1), 130--155.

\bibitem[\protect\citeauthoryear{Ing and Wei}{Ing and Wei}{2005}]{ing2005order}
Ing, C.-K. and C.-Z. Wei (2005).
\newblock Order selection for same-realization predictions in autoregressive
  processes.
\newblock {\em The Annals of Statistics\/}~{\em 33\/}(5), 2423--2474.

\bibitem[\protect\citeauthoryear{Kabaila}{Kabaila}{1999}]{kabaila1999relevance}
Kabaila, P. (1999).
\newblock The relevance property for prediction intervals.
\newblock {\em Journal of Time Series Analysis\/}~{\em 20\/}(6), 655--662.

\bibitem[\protect\citeauthoryear{Kabaila and He}{Kabaila and
  He}{2004}]{kabaila2004adjustment}
Kabaila, P. and Z.~He (2004).
\newblock The adjustment of prediction intervals to account for errors in
  parameter estimation.
\newblock {\em Journal of Time Series Analysis\/}~{\em 25\/}(3), 351--358.

\bibitem[\protect\citeauthoryear{Kabaila and Syuhada}{Kabaila and
  Syuhada}{2008}]{kabaila2008improved}
Kabaila, P. and K.~Syuhada (2008).
\newblock Improved prediction limits for \uppercase{AR}($p$) and
  \uppercase{ARCH}($p$) processes.
\newblock {\em Journal of Time Series Analysis\/}~{\em 29\/}(2), 213--223.

\bibitem[\protect\citeauthoryear{Kabaila and Syuhada}{Kabaila and
  Syuhada}{2010}]{kabaila2010asymptotic}
Kabaila, P. and K.~Syuhada (2010).
\newblock The asymptotic efficiency of improved prediction intervals.
\newblock {\em Statistics \& Probability Letters\/}~{\em 80\/}(17), 1348--1353.

\bibitem[\protect\citeauthoryear{Kreiss}{Kreiss}{2016}]{kreiss2015discussion}
Kreiss, J.-P. (2016).
\newblock Discussion: bootstrap prediction intervals for linear, nonlinear and
  nonparametric autoregressions.
\newblock {\em Journal of Statistical Planning and Inference\/}~{\em 177},
  28--30.

\bibitem[\protect\citeauthoryear{Kunitomo and Yamamoto}{Kunitomo and
  Yamamoto}{1985}]{kunitomo1985properties}
Kunitomo, N. and T.~Yamamoto (1985).
\newblock Properties of predictors in misspecified autoregressive time series
  models.
\newblock {\em Journal of the American Statistical Association\/}~{\em
  80\/}(392), 941--950.

\bibitem[\protect\citeauthoryear{Lewis and Reinsel}{Lewis and
  Reinsel}{1985}]{lewis1985prediction}
Lewis, R. and G.~C. Reinsel (1985).
\newblock Prediction of multivariate time series by autoregressive model
  fitting.
\newblock {\em Journal of Multivariate Analysis\/}~{\em 16\/}(3), 393--411.

\bibitem[\protect\citeauthoryear{L{\"u}tkepohl}{L{\"u}tkepohl}{2005}]{lutkepohl2005new}
L{\"u}tkepohl, H. (2005).
\newblock {\em New Introduction to Multiple Time Series Analysis}.
\newblock Berlin: Springer.

\bibitem[\protect\citeauthoryear{Pan and Politis}{Pan and
  Politis}{2016a}]{pan2016abootstrap}
Pan, L. and D.~N. Politis (2016a).
\newblock Bootstrap prediction intervals for linear, nonlinear and
  nonparametric autoregressions.
\newblock {\em Journal of Statistical Planning and Inference\/}~{\em 177},
  1--27.

\bibitem[\protect\citeauthoryear{Pan and Politis}{Pan and
  Politis}{2016b}]{pan2016bbootstrap}
Pan, L. and D.~N. Politis (2016b).
\newblock Bootstrap prediction intervals for \uppercase{M}arkov processes.
\newblock {\em Computational Statistics \& Data Analysis\/}~{\em 100},
  467--494.

\bibitem[\protect\citeauthoryear{Pascual, Romo, and Ruiz}{Pascual
  et~al.}{2004}]{pascual2004bootstrap}
Pascual, L., J.~Romo, and E.~Ruiz (2004).
\newblock Bootstrap predictive inference for \uppercase{ARIMA} processes.
\newblock {\em Journal of Time Series Analysis\/}~{\em 25\/}(4), 449--465.

\bibitem[\protect\citeauthoryear{Pascual, Romo, and Ruiz}{Pascual
  et~al.}{2006}]{pascual2006bootstrap}
Pascual, L., J.~Romo, and E.~Ruiz (2006).
\newblock Bootstrap prediction for returns and volatilities in
  \uppercase{garch} models.
\newblock {\em Computational Statistics \& Data Analysis\/}~{\em 50\/}(9),
  2293--2312.

\bibitem[\protect\citeauthoryear{Pesaran}{Pesaran}{2015}]{pesaran2015time}
Pesaran, M.~H. (2015).
\newblock {\em Time Series and Panel Data Econometrics}.
\newblock Oxford: Oxford University Press.

\bibitem[\protect\citeauthoryear{Phillips}{Phillips}{1979}]{phillips1979sampling}
Phillips, P. C.~B. (1979).
\newblock The sampling distribution of forecasts from a first-order
  autoregression.
\newblock {\em Journal of Econometrics\/}~{\em 9\/}(3), 241--261.

\bibitem[\protect\citeauthoryear{Samaranayake and Hasza}{Samaranayake and
  Hasza}{1988}]{samaranayake1988properties}
Samaranayake, V.~A. and D.~P. Hasza (1988).
\newblock Properties of predictors for multivariate autoregressive models with
  estimated parameters.
\newblock {\em Journal of Time Series Analysis\/}~{\em 9\/}(4), 361--383.

\bibitem[\protect\citeauthoryear{van~der Vaart}{van~der
  Vaart}{2000}]{van2000asymptotic}
van~der Vaart, A.~W. (2000).
\newblock {\em Asymptotic Statistics}.
\newblock Cambridge: Cambridge University Press.

\bibitem[\protect\citeauthoryear{Vidoni}{Vidoni}{2004}]{vidoni2004improved}
Vidoni, P. (2004).
\newblock Improved prediction intervals for stochastic process models.
\newblock {\em Journal of Time Series Analysis\/}~{\em 25\/}(1), 137--154.

\bibitem[\protect\citeauthoryear{Vidoni}{Vidoni}{2009a}]{vidoni2009improved}
Vidoni, P. (2009a).
\newblock Improved prediction intervals and distribution functions.
\newblock {\em Scandinavian Journal of Statistics\/}~{\em 36\/}(4), 735--748.

\bibitem[\protect\citeauthoryear{Vidoni}{Vidoni}{2009b}]{vidoni2009simple}
Vidoni, P. (2009b).
\newblock A simple procedure for computing improved prediction intervals for
  autoregressive models.
\newblock {\em Journal of Time Series Analysis\/}~{\em 30\/}(6), 577--590.

\bibitem[\protect\citeauthoryear{Vidoni}{Vidoni}{2017}]{vidoni2016improved}
Vidoni, P. (2017).
\newblock Improved multivariate prediction regions for \uppercase{M}arkov
  process models.
\newblock {\em Statistical Methods \& Applications\/}~{\em 26\/}(1), 1--18.

\bibitem[\protect\citeauthoryear{Xiong and Li}{Xiong and
  Li}{2008}]{xiong2008some}
Xiong, S. and G.~Li (2008).
\newblock Some results on the convergence of conditional distributions.
\newblock {\em Statistics \& Probability Letters\/}~{\em 78\/}(18), 3249--3253.

\bibitem[\protect\citeauthoryear{Zako{\"\i}an}{Zako{\"\i}an}{1994}]{zakoian1994threshold}
Zako{\"\i}an, J.-M. (1994).
\newblock Threshold heteroskedastic models.
\newblock {\em Journal of Economic Dynamics and Control\/}~{\em 18\/}(5),
  931--955.

\end{thebibliography}

%%%%%%%%%%%%%%%%%%%%
%%%%%APPENDIX A%%%%%
%%%%%%%%%%%%%%%%%%%%

\appendix

\section{Lemmas and Proofs of Theorems}
\label{app:A}

We only present the proofs of the leading results here. The proofs of the lemmas and corollaries can be found in Appendix \ref{app:B}. Before going to the proofs, we first introduce the following auxiliary metrics that will be encountered in the proofs. For arbitrary cdfs $F$ and $G$ on $\R$, the Kolmogorov and  L\'evy metric are
\begin{align*}
d_K(F,G)=&\sup_{\tau \in \R}\big|F(\tau)-G(\tau)\big|\\
d_L(F,G)=&\inf\big\{\xi>0: G(\tau-\xi)-\xi\leq F(\tau)\leq G(\tau+\xi)+\xi \quad \forall \tau \in \R\big\}\:.
\end{align*}
Moreover, let $Z_{\infty}\sim G_{\infty}$ (with $G_{\infty}$ given in Assumption \ref{as:2.0}.a) be defined on some probability space $(\breve{\Omega},\breve{\mathcal{F}},\breve{\PP})$ and define the product measure $\bar{\PP}=\PP \times \breve{\PP}$ on the space $\Omega \times \breve{\Omega} $ with $\sigma$-field, generated by the \textit{measurable rectangles} $A\times \breve{A}$ with $A\in \mathcal{F}$ and $\breve{A}\in \breve{\mathcal{F}}$ (cf.\ \citeauthor{billingsley1986probability}, \citeyear{billingsley1986probability}, Thm.\ 18.2). Notice that $Z_{\infty}$ is independent of $\{X_t\}$ and  $\{Y_t\}$ by construction.

In the proofs we often consider the ``doubly random'' versions of estimators and intervals, where the first function argument, the sample to condition on, is considered random. Instead, we account for the conditioning in the probability statements. This notation is more convenient for proving the results, as we need them to hold for ``all sequences $\x_{1:T}$ occurring with high probability'', which is much easier to quantify by treating these sequences as random. Therefore we use notations such as the unconditional estimators $\hat\psi_{T+1}^{2IP} = \hat\psi_{T+1} (\X_{1:T}, \Y_{1:T})$ and $\hat\psi_{T+1}^{SPL} = \hat\psi_{T+1} (\X_{T_P:T}^c, \X_{1:T_E})$, as well as their corresponding intervals $I^{2IP}(\X_{1:T},\Y_{1:T})$ and $I^{SPL}(\X_{T_P:T}^c,\X_{1:T_E})$; also see (\ref{eq:2.4.2})-(\ref{eq:2.4.3}) and the remarks above Theorem \ref{thm:2.3}.

\subsection{Lemmas}

\begin{lemma}
\label{lem:2.1}
Let
\begin{small}
\begin{align}
\label{eq:2.4.4}
R_T^{2IP} &:= m_T\Big(\hat{\psi}_{T+1} (\X_{1:T}, \Y_{1:T}) - \psi_{T+1}\Big) - \frac{\partial \psi_{T+1}^s(\X_{1:T}, \theta_0)}{\partial \theta'} m_T \big(\hat{\theta}(\Y_{1:T}) - \theta_0\big)\\
\label{eq:2.4.5}
R_T^{SPL} &:= m_T\Big(\hat{\psi}_{T+1} (\X_{T_P:T}^c, \X_{1:T_E})  - \psi_{T+1}\Big) - \frac{\partial \psi_{T+1}^s (\X_{T_P:T}^c;\theta_0)}{\partial \theta'} m_T\big(\hat{\theta}(\X_{1:T_E}) - \theta_0\big).
\end{align}
\end{small}
\begin{enumerate}[(i)]
\item If Assumptions \ref{as:2.0}.a, \ref{as:2.0}.b, \ref{as:2.0}.d, \ref{as:2.0}.e and \ref{as:2.1}.a hold, then $R_T^{2IP}$  is $o_{p}(1)$;
\item if Assumptions \ref{as:2.0}.a, \ref{as:2.0}.b, \ref{as:2.0}.d, \ref{as:2.0}.e and \ref{as:2.2}.a hold, then $R_T^{SPL}$ is $o_{p}(1)$.
\end{enumerate}
\end{lemma}

\begin{lemma}
\label{lem:2.2}
Let $G_{T_E}^{SPL}(\cdot|\mathcal{I}_{T_P:T})$ be as given in Assumption \ref{as:2.2}.c and denote the conditional cdf of $Z_T^{2IP}:=m_{T}\big(\hat{\theta}(\mathbf{Y}_{1:T}) - \theta_0\big)$  given $\mathcal{I}_{1:T}$ by $G_{T}^{2IP}(\cdot|\mathcal{I}_{1:T})$.
\begin{enumerate}[(i)]
\item Under Assumptions \ref{as:2.0}.a and \ref{as:2.1}, $\int h \: dG_{T}^{2IP}(\cdot|\mathcal{I}_{1:T})\overset{p}{\to}\int h \: d G_{\infty}$ $\forall h \in BL$;
\item Under Assumptions \ref{as:2.0}.a and \ref{as:2.2}, $\int h \: d G_{T_E}^{SPL}(\cdot|\mathcal{I}_{T_P:T})\overset{p}{\to}\int h \: d G_{\infty}$  $\forall h \in BL$.
\end{enumerate}
\end{lemma}
\begin{lemma}
\label{lem:2.3}
Let $\{G_T\}$ be a sequence of cdfs and $G$ be a (non-random) cdf on $(\R^r,||\cdot||)$. If $\int h\:dG_T \overset{p}{\to}\int h\:dG$ for all $h \in BL$, then $\sup_{h \in \mathscr{H}} \big|\int h\: d(G_T-G)\big|\overset{p}{\to}0$, where $\mathscr{H}=\big\{h:\R^r\to \R:||h||_{BL}\leq 1\big\}$.
\end{lemma}
\begin{lemma}
\label{lem:2.4}
Assume that the $\R^r$-valued random variable $w_T$ is $O_p(1)$ and $\mathcal{I}_T$-measurable. Further, suppose the real-valued random variable $R_T$ is $o_p(1)$ and the  $\R^r$-valued random variable $Z_T$ satisfies $\bar{\PP}[Z_T \leq \cdot|\mathcal{I}_T]\overset{p}{\to} G_{\infty}$. Then, the two sequences of conditional cdfs $\bar{\PP}[w_T'Z_T+R_T\leq \cdot|\mathcal{I}_T]$ and $\bar{\PP}[w_T'Z_{\infty} \leq \cdot|\mathcal{I}_T]$ merge in probability.
\end{lemma}
\begin{lemma}
\label{lem:2.5}
Let $\epsilon>0$ and  $F$ and $G$ be cdfs on $\R$ with $G(\tau-\epsilon)-\epsilon \leq F(\tau)\leq G(\tau+\epsilon)+\epsilon$ for all $\tau \in \R$. Then $F^{-1}(u-\epsilon)-\epsilon \leq G^{-1}(u)\leq F^{-1}(u+\epsilon)+\epsilon$ for all $u \in (\epsilon,1-\epsilon)$.
\end{lemma}
\begin{lemma}
\label{lem:2.6}
Suppose $\{F_T\}$ and $\{G_T\}$ are sequences of conditional cdfs with $d_L(F_T,G_T)\overset{p}{\to} 0$ as $T\to \infty$.
Further, assume that $G_T$ is stochastically uniformly equicontinuous: for every $\epsilon,\eta >0$, there exist $\delta=\delta(\epsilon,\eta )>0$  and $\bar{T}=\bar{T}(\epsilon,\eta)$ such that $\PP\big[\sup\limits_{\tau \in \R}\sup\limits_{\tau'\in\R:|\tau-\tau'|<\delta} |G_T(\tau')-G_T(\tau)|>\epsilon\big]<\eta $ for all $T\geq \bar{T}$. Then, $d_K(F_T,G_T)\overset{p}{\to} 0$.
\end{lemma}
\begin{lemma}
\label{lem:2.7}
If the sequences of conditional cdfs $\{F_T\}$ and $\{G_T\}$ merge in probability and $G_T$ is stochastically uniformly equicontinuous, then $\PP\big[G_T^{-1}(\gamma_1) \leq M_T \leq G_T^{-1}(1-\gamma_2) \big|\mathcal{I}_T\big]\overset{p}{\to}1-\gamma_1-\gamma_2$ whenever $0\leq \gamma_1\leq 1-\gamma_2\leq 1$, where the random variable $M_T$ given the $\sigma$-algebra $\mathcal{I}_T$ has the cdf $F_T$.
\end{lemma}

\begin{lemma}
\label{lem:2.8}
Suppose $\{F_T\}$ and $\{G_T\}$ are sequences of conditional cdfs with $d_L(F_T,G_T)\overset{p}{\to} 0$ as $T\to \infty$. Further, assume that $G_T^{-1}$ is stochastically pointwise equicontinuous at $u\in(0,1)$: for all $\epsilon,\eta >0$, there exist  $\delta=\delta(\epsilon,\eta ,u)>0$ and  $\bar{T}=\bar{T}(\epsilon,\eta ,u)$ such that $\PP\Big[\sup\limits_{|u-v|<\delta}\big|G_T^{-1}(v)-G_T^{-1}(u)\big|>\epsilon\Big]<\eta $ for all $T\geq \bar{T}$. Then $\big|F_T^{-1}(u)-G_T^{-1}(u)\big|\overset{p}{\to}0$.
\end{lemma}

\subsection{Proofs of Theorems.}

\noindent
\textbf{Proof of Theorem \ref{thm:2.1}.}
Let $w_T^{2IP}$ equal the transpose of $\frac{\partial \psi_{T+1}^s(\mathbf{X}_{1:T};\theta_0)}{\partial \theta}$ and $w_T^{SPL}$ the transpose of $\frac{\partial \psi_{T+1}^s(\mathbf{X}_{T_P:T}^c;\theta_0)}{\partial \theta}$ and set
\begin{align}
\label{eq:2.A.1}
F_{\infty,T} ^{2IP}(\tau|\mathcal{I}_{1:T}):=&\bar{\PP}\big[w_T^{2IP} Z_{\infty} \leq \tau \big|\mathcal{I}_{1:T}\big]\\
\label{eq:2.A.2}
F_{\infty,T} ^{SPL}(\tau|\mathcal{I}_{T_P:T}):=&\bar{\PP}\big[w_T^{SPL}Z_{\infty} \leq \tau \big|\mathcal{I}_{T_P:T}\big]\:,
\end{align}
the conditional cdfs of $w_T^{2IP}Z_{\infty}$ given $\mathcal{I}_{1:T}$ and $w_T^{SPL}Z_{\infty}$ given $\mathcal{I}_{T_P:T}$, respectively. Note that (\ref{eq:2.A.1}) and (\ref{eq:2.A.2}) can be considered to be the 'merging limits' of $w^{2IP}_Tm_T(\hat{\theta}(\mathbf{X}_{1:T})-\theta_0)$ and $w^{SPL}_T m_{T_E}(\hat{\theta}(\mathbf{X}_{1:T_E})-\theta_0)$. Since $m_T(\hat{\theta}(\mathbf{X}_{1:T})-\theta_0)$ and $m_{T_E}(\hat{\theta}(\mathbf{X}_{1:T_E})-\theta_0)$ converge in distribution to $G_{\infty}$ but $w^{2IP}_T$ and $w_T^{SPL}$ do not converge, we indexed the 'merging limits' in (\ref{eq:2.A.1}) and (\ref{eq:2.A.2}) by $\infty$ and $T$. The triangle inequality implies
\begin{align*}
&d_{BL}\Big(F_T^{2IP}(\cdot|\mathcal{I}_{1:T}),F_T^{SPL}(\cdot|\mathcal{I}_{T_P:T})\Big)\leq \underbrace{d_{BL}\Big(F_T^{2IP}(\cdot|\mathcal{I}_{1:T}),F_{\infty,T}^{2IP}(\cdot|\mathcal{I}_{1:T})\Big)}_{I}\\
&\qquad+\underbrace{d_{BL}\Big(F_{\infty,T}^{2IP}(\cdot|\mathcal{I}_{1:T}),F_{\infty,T} ^{SPL}(\cdot|\mathcal{I}_{T_P:T})\Big)}_{II}
+\underbrace{d_{BL}\Big(F_{\infty,T} ^{SPL}(\cdot|\mathcal{I}_{T_P:T}),F_T^{SPL}(\cdot|\mathcal{I}_{T_P:T}\Big)}_{III}.
\end{align*}
We prove that $I$, $II$ and $III$ are $o_p(1)$. Consider $I$ and note that $m_T\big(\hat{\psi}_{T+1}^{2IP}-\psi_{T+1}\big)=w_T^{2IP}Z_T^{2IP}+R_T^{2IP}$, where $Z_T^{2IP}=m_T\big(\hat{\theta}(\Y_{1:T})-\theta_0\big)$. The weight $w_T^{2IP}$ is $\mathcal{I}_{1:T}$ measurable and $O_p(1)$ by Assumptions \ref{as:2.0}.c and \ref{as:2.0}.e, $R_T^{2IP}$ is $o_p(1)$ by Lemma \ref{lem:2.1} and $\int h\: dG_{T}^{2IP}(\cdot|\mathcal{I}_{1:T})\overset{p}{\to}\int h\: dG_{\infty}$ for each $h \in BL$ by Lemma \ref{lem:2.2}. Replacing $Z_T$, $R_T$, $w_T$ and $\mathcal{I}_T$ in Lemma \ref{lem:2.4} by $Z_T^{2IP}$, $R_T^{2IP}$, $w_T^{2IP}$ and $\mathcal{I}_{1:T}$ implies that $F_T^{2IP}(\tau|\mathcal{I}_{1:T})=\bar{\PP}[w_T^{2IP}Z_T^{2IP}+R_T^{2IP}\leq \tau|\mathcal{I}_{1:T}]$ and $F_{\infty,T}^{2IP}(\tau|\mathcal{I}_{1:T})=\bar{\PP}[w_T^{2IP}Z_{\infty} \leq \tau|\mathcal{I}_{1:T}]$ merge in probability, i.e.\ $I\overset{p}{\to}0$.

Consider $II$ and rewrite $w_T^{SPL}Z_{\infty}=w_T^{2IP}Z_{\infty}+(w_T^{SPL}-w_T^{2IP})Z_{\infty}$. The weight $w_T^{2IP}$ is $\mathcal{I}_{1:T}$ measurable and $O_p(1)$ by Assumptions \ref{as:2.0}.c and \ref{as:2.0}.e, $(w_T^{SPL}-w_T^{2IP})Z_{\infty}=o_p(1)\:O_p(1)=o_p(1)$ by Assumption \ref{as:2.0}.e and $\int h \:d\bar{\PP}[Z_{\infty} \leq \cdot|\mathcal{I}_{1:T}]=\int h\:d G_{\infty}$ for each $h \in BL$. Replacing $Z_T$, $R_T$, $w_T$ and $\mathcal{I}_T$ in Lemma \ref{lem:2.4} by $Z_{\infty}$, $(w_T^{SPL}-w_T^{2IP})Z_{\infty}$, $w_T^{2IP}$ and $\mathcal{I}_{1:T}$ implies that $F_{\infty,T}^{2IP}(\tau|\mathcal{I}_{1:T})=\bar{\PP}[w_T^{2IP}Z_{\infty} \leq \tau|\mathcal{I}_{1:T}]$ and $F_{\infty,T} ^{SPL}(\tau|\mathcal{I}_{T_p:T})=\bar{\PP}[w_T^{SPL}Z_{\infty}\leq \tau|\mathcal{I}_{1:T}]$ merge in probability,  i.e.\ $II\overset{p}{\to}0$.

Consider $III$ and note that $m_T\big(\hat{\psi}_{T+1}^{SPL}-\psi_{T+1}\big)=w_T^{SPL}Z_T^{SPL}+R_T^{SPL}+S_T^{SPL}$, where $Z_T^{SPL}=m_{T_E}\big(\hat{\theta}(\X_{1:T_E})-\theta_0\big)$ and $S_T^{SPL}=\big(\frac{m_T}{m_{T_E}}-1\big)w_T^{SPL}Z_T^{SPL}$. The weight $w_T^{SPL}$ is $\mathcal{I}_{T_P:T}$ measurable and $O_p(1)$ by Assumptions \ref{as:2.0}.c and \ref{as:2.0}.e, $R_T^{SPL}$ is $o_p(1)$ by Lemma \ref{lem:2.1} and $\int h\: dG_{T_E}^{SPL}(\cdot|\mathcal{I}_{T_P:T})\overset{p}{\to}\int h\: d G_{\infty}$ for each $h \in BL$  by Lemma \ref{lem:2.2}. Further, $S_T^{SPL}$ is $o_p(1)$ since $w_T^{SPL}=O_p(1)$ and $\big(\frac{m_T}{m_{T_E}}-1\big)Z_T^{SPL}=o_p(1)$ by Assumptions \ref{as:2.0}.a and \ref{as:2.2}.a. Replacing $Z_T$, $R_T$, $w_T$ and $\mathcal{I}_T$ in Lemma \ref{lem:2.4} by $Z_T^{SPL}$, $R_T^{SPL}+S_T^{SPL}$, $w_T^{SPL}$ and $\mathcal{I}_{T_P:T}$ implies that $F_T^{SPL}(\tau|\mathcal{I}_{T_P:T})=\bar{\PP}[w_T^{SPL}Z_T^{SPL}+R_T^{SPL}+S_T^{SPL}\leq \tau|\mathcal{I}_{T_P:T}]$ and $F_{\infty,T} ^{SPL}(\tau|\mathcal{I}_{T_P:T})=\bar{\PP}[w_T^{SPL}Z_{\infty} \leq \tau|\mathcal{I}_{T_P:T}]$ merge in probability,  i.e.\ $III\overset{p}{\to}0$. \qed \\ \\
\noindent
\textbf{Proof of Theorem \ref{thm:2.2}.}
Consider statement 1(a). Let $\hat{Z}_T^{2IP}$ follow the mixture distribution $\widehat{G}_T (\cdot)$ as a function of $\Y_{1:T}$ such that given $\Y_{1:T}$ the conditional distribution of the random variable $\hat{Z}_T^{2IP}$ is $\widehat{G}_T (\cdot)$. Further, let 
\begin{align}\label{eq:2.A.3}
\mbox{$\widehat{F_T^{2IP}}(\cdot)$ be the conditional cdf of $\hat{w}_T^{2IP}\hat{Z}_T^{2IP}$ given $\mathcal{H}_{1:T}$}
\end{align}
where $\hat{w}_T^{2IP}$ equals the transpose of $\frac{\partial \psi_T^s(\mathbf{X}_{1:T};\hat{\theta}(\mathbf{Y}_{1:T}))}{\partial \theta}$ and $\mathcal{H}_{1:T}=\sigma\big(X_1,\dots,X_T,Y_1,\dots,Y_T\big)$. Then
\begin{align*}
&d_{BL}\Big(F_T^{2IP}(\cdot|\mathcal{I}_{1:T}),\widehat{F_T^{2IP}}(\cdot)\Big)
\leq d_{BL}\Big(F_T^{2IP}(\cdot|\mathcal{I}_{1:T}),F_{\infty,T}^{2IP}(\cdot|\mathcal{I}_{1:T})\Big)+d_{BL}\Big(F_{\infty,T}^{2IP}(\cdot|\mathcal{I}_{1:T}),\widehat{F_T^{2IP}}(\cdot)\Big)
\end{align*}
by the triangle inequality, where $F_{\infty,T}^{2IP}(\cdot|\mathcal{I}_{1:T})$ is defined in equation \eqref{eq:2.A.1}. In the proof of Theorem \ref{thm:2.1}, we have shown that $F_T^{2IP}(\cdot|\mathcal{I}_{1:T})$ and $F_{\infty,T}^{2IP}(\cdot|\mathcal{I}_{1:T})$ merge in probability under Assumptions \ref{as:2.0} and \ref{as:2.1}. It suffices to show that $F_{\infty,T}^{2IP}(\cdot|\mathcal{I}_{1:T})$ and $\widehat{F_T^{2IP}}(\cdot)$ merge in probability. Write $\hat{w}_T^{2IP}\hat{Z}_T^{2IP}=w_T^{2IP}\hat{Z}_T^{2IP}+\hat{R}_T^{2IP}$ with $\hat{R}_T^{2IP}= (\hat{w}_T^{2IP}-w_T^{2IP})\hat{Z}_T^{2IP}$. First, we show $\hat{R}_T^{2IP}=o_p(1)$. Take an arbitrary $\epsilon>0$. We obtain
\begin{align*}
&\PP \Big[\big|\hat{R}_T^{2IP}\big|\geq\epsilon\Big]
\leq \PP \Bigg[\bigg|\bigg|\frac{\partial^2 \psi_{T+1}^s(\mathbf{X}_{1:T};\dot{\theta}_T)}{\partial \theta \partial \theta'}\bigg|\bigg| \Big|\Big| \hat{\theta}(\mathbf{Y}_{1:T}) - \theta_0\Big|\Big|\: \big|\big|\hat{Z}_T^{2IP}\big|\big|\geq\epsilon\Bigg]\\
&\leq \PP \Bigg[\bigg|\bigg|\frac{\partial^2 \psi_{T+1}^s(\mathbf{X}_{1:T};\dot{\theta}_T)}{\partial \theta \partial \theta'}\bigg|\bigg| \:\Big|\Big| \hat{\theta}(\mathbf{Y}_{1:T}) - \theta_0\Big|\Big|\: \big|\big|\hat{Z}_T^{2IP}\big|\big|\geq\epsilon \bigcap \dot{\theta}_T \in \mathscr{V}(\theta_0)\Bigg] + \PP\Big[ \dot{\theta}_T \notin \mathscr{V}(\theta_0)\Big]\\
&\leq \PP \Bigg[\sup_{\theta \in  \mathscr{V}(\theta_0)}\bigg|\bigg|\frac{\partial^2 \psi_{T+1}^s(\mathbf{X}_{1:T};\theta)}{\partial \theta \partial \theta'}\bigg|\bigg| \:\Big|\Big| \hat{\theta}(\mathbf{Y}_{1:T}) - \theta_0\Big|\Big|\: \big|\big|\hat{Z}_T^{2IP}\big|\big|\geq\epsilon \Bigg] + \PP\Big[ \dot{\theta}_T \notin \mathscr{V}(\theta_0)\Big]\:,
\end{align*}
where $\dot{\theta}_T$ lies between $\hat{\theta}(\mathbf{Y}_{1:T})$ and $\theta_0$. The first term vanishes as $\sup\limits_{\theta \in \mathscr{V}(\theta_0)}\Big|\Big|\frac{\partial^2 \psi_{T+1}(\mathbf{X}_{1:T};\theta)}{\partial \theta \partial \theta'}\Big|\Big|=O_{p}(1)$ by Assumptions \ref{as:2.0}.d and \ref{as:2.0}.e,  $\big|\big|\hat{\theta}(\mathbf{Y}_{1:T}) - \theta_0\big|\big|=O_p(m_T^{-1})$ by Assumptions \ref{as:2.0}.a and \ref{as:2.1}.a and $\big|\big|\hat{Z}_T^{2IP}\big|\big|=O_p(1)$ as $\hat{Z}_T^{2IP}\sim \widehat{G}_T (\cdot)$ and $\int h \:d \widehat{G}_T\overset{p}{\to}\int h \:d G_{\infty}$ for all $h \in BL$ by Assumptions \ref{as:2.1}.a and \ref{as:2.4}. Further, as $\hat{\theta}(\mathbf{Y}_{1:T})\overset{p}{\to} \theta_0\in\mathscr{V}(\theta_0)$ and $\mathscr{V}(\theta_0)$ is open, we have  $\PP\big[ \dot{\theta}_T \notin \mathscr{V}(\theta_0)\big]\to 0$ and $\hat{R}_T^{2IP}=o_p(1)$ follows. Moreover, $w_T^{2IP}$ is $\mathcal{H}_T$-measurable and $O_p(1)$ by Assumptions \ref{as:2.0}.c and \ref{as:2.0}.e and $\int h \:d\bar{\PP}\big[\hat{Z}_T^{2IP}\leq \cdot|\mathcal{H}_{1:T}\big]=\int h \:d\widehat{G}_T \overset{p}{\to}\int h \: dG_{\infty}$ for each $h \in BL$. Replacing $Z_T$, $R_T$, $w_T$ and $\mathcal{I}_T$ in Lemma \ref{lem:2.4} by $\hat{Z}_T^{2IP}$, $\hat{R}_T^{2IP}$, $w_T^{2IP}$ and $\mathcal{H}_T$ implies that $\widehat{F_T^{2IP}}(\cdot)=\bar{\PP}[w_T^{2IP} \hat{Z}_T^{2IP}+\hat{R}_T^{2IP}\leq \cdot|\mathcal{H}_{1:T}]$ and $\bar{\PP}[w_T^{2IP}Z_{\infty} \leq \cdot|\mathcal{H}_{1:T}]=F_{\infty,T}^{2IP}(\cdot|\mathcal{I}_{1:T})$ merge in probability.

Consider statement 1(b). As $F_T^{2IP}(\cdot|\mathcal{I}_{1:T})$ and $\widehat{F_T^{2IP}}(\cdot)$ merge in probability and $\widehat{F_T^{2IP}}(\cdot)$ is assumed to be stochastically uniformly continuous, Lemma \ref{lem:2.7} applies. Replacing $F_T$, $G_T$, $M_T$ and $\mathcal{I}_T$ by $F_T^{2IP}(\cdot|\mathcal{I}_{1:T})$, $\widehat{F_T^{2IP}}(\cdot)$, $m_T\big(\hat{\psi}_{T+1}^{2IP}- \psi_{T+1} \big)$ and $\mathcal{I}_{1:T}$, respectively, it follows that
\begin{align*}
&\PP \Big[  I_\gamma^{2IP}(\X_{1:T},\Y_{1:T}) \ni \psi_{T+1} \Big|\mathcal{I}_{1:T}\Big]\\
&=\PP \Big[ \widehat{F_T^{2IP}}^{-1}(\gamma_1)\leq m_T\big(\hat{\psi}_{T+1}^{2IP}- \psi_{T+1} \big) \leq \widehat{F_T^{2IP}}^{-1}(1-\gamma_2) \Big|\mathcal{I}_{1:T}\Big]\overset{p}{\to} 1 -\gamma\:.
\end{align*}

Claim 2(a) is similarly proven as 1(a). Let $\hat{Z}_T^{SPL}$ follow the mixture distribution $\widehat{G}_T (\cdot)$ as a function of $\X_{1:T_E}$ such that given $\X_{1:T_E}$ the conditional distribution of the random variable $\hat{Z}_T^{SPL}$ is $\widehat{G}_T (\cdot)$. Further, let 
\begin{align}\label{eq:2.A.4}
\mbox{$\widehat{F_T^{SPL}}(\cdot)$ be the conditional cdf of $\hat{w}_T^{SPL}\hat{Z}_T^{SPL}$ given $\mathcal{I}_{1:T}$}
\end{align}
where $\hat{w}_T^{SPL}$ equals the transpose of $\frac{\partial \psi_T^s(\mathbf{X}_{T_P:T}^c;\hat{\theta}(\X_{1:T_E}))}{\partial \theta}$. Then
\begin{align*}
&d_{BL}\Big(F_T^{SPL}(\cdot|\mathcal{I}_{T_P:T}),\widehat{F_T^{SPL}}(\cdot)\Big)\\
&\leq d_{BL}\Big(F_T^{SPL}(\cdot|\mathcal{I}_{T_P:T}),F_{\infty,T} ^{SPL}(\cdot|\mathcal{I}_{T_P:T})\Big)+d_{BL}\Big(F_{\infty,T} ^{SPL}(\cdot|\mathcal{I}_{T_P:T}),\widehat{F_T^{SPL}}(\cdot)\Big),
\end{align*}
where $F_{\infty,T} ^{SPL}(\cdot|\mathcal{I}_{T_P:T})$ is defined in equation \eqref{eq:2.A.2}. In the proof of Theorem \ref{thm:2.1}, we have shown that $F_T^{SPL}(\cdot|\mathcal{I}_{T_P:T})$ and $F_{\infty,T} ^{SPL}(\cdot|\mathcal{I}_{T_P:T})$ merge in probability under Assumptions \ref{as:2.0} and \ref{as:2.2}. It suffices to show that $F_{\infty,T} ^{SPL}(\cdot|\mathcal{I}_{T_P:T})$ and $\widehat{F_T^{SPL}}(\cdot)$ merge in probability. Write $\hat{w}_T^{SPL}\hat{Z}_T^{SPL}=w_T^{SPL}\hat{Z}_T^{SPL}+\hat{R}_T^{SPL}$ with $\hat{R}_T^{SPL}= (\hat{w}_T^{SPL}-w_T^{SPL})\hat{Z}_T^{SPL}$. First, we show $\hat{R}_T^{SPL}=o_p(1)$. Take an arbitrary $\epsilon>0$. We obtain
\begin{align*}
&\PP \Big[\big|\hat{R}_T^{SPL}\big|\geq\epsilon\Big]\leq \PP \Bigg[\bigg|\bigg|\frac{\partial^2 \psi_{T+1}^s(\X_{T_P:T}^c;\ddot{\theta}_T)}{\partial \theta \partial \theta'}\bigg|\bigg|\: \Big|\Big| \hat{\theta}(\X_{1:T_E}) - \theta_0\Big|\Big|\: \big|\big|\hat{Z}_T^{SPL}\big|\big|\geq\epsilon\Bigg]\\
&\leq \PP \Bigg[\bigg|\bigg|\frac{\partial^2 \psi_{T+1}^s(\X_{T_P:T}^c;\ddot{\theta}_T)}{\partial \theta \partial \theta'}\bigg|\bigg| \:\Big|\Big| \hat{\theta}(\X_{1:T_E}) - \theta_0\Big|\Big|\: \big|\big|\hat{Z}_T^{SPL}\big|\big|\geq\epsilon \bigcap \ddot{\theta}_T \in \mathscr{V}(\theta_0)\Bigg] + \PP\Big[ \ddot{\theta}_T \notin \mathscr{V}(\theta_0)\Big]\\
&\leq \PP \Bigg[\sup_{\theta \in  \mathscr{V}(\theta_0)}\bigg|\bigg|\frac{\partial^2 \psi_{T+1}^s(\X_{T_P:T}^c;\theta)}{\partial \theta \partial \theta'}\bigg|\bigg| \:\Big|\Big| \hat{\theta}(\X_{1:T_E}) - \theta_0\Big|\Big|\: \big|\big|\hat{Z}_T^{SPL}\big|\big|\geq\epsilon \Bigg] + \PP\Big[ \ddot{\theta}_T \notin \mathscr{V}(\theta_0)\Big]\:,
\end{align*}
where $\ddot{\theta}_T$ lies between $\hat{\theta}(\X_{1:T_E})$ and $\theta_0$. The first term vanishes as $\sup\limits_{\theta \in \mathscr{V}(\theta_0)}\Big|\Big|\frac{\partial^2 \psi_{T+1}^s(\X_{T_P:T}^c;\theta)}{\partial \theta \partial \theta'}\Big|\Big|$ is $O_{p}(1)$ by Assumptions \ref{as:2.0}.d and \ref{as:2.0}.e and  $\big|\big|\hat{\theta}(\X_{1:T_E}) - \theta_0\big|\big|=O_p(m_{T_E}^{-1})$ by Assumptions \ref{as:2.0}.a and \ref{as:2.2}.a, $\big|\big|\hat{Z}_T^{SPL}\big|\big|=O_p(1)$ as $\hat{Z}_T^{SPL}\sim \widehat{G}_T (\cdot)$ and $\int h \:d \widehat{G}_T\overset{p}{\to}\int h \:d G_{\infty}$ for all $h \in BL$ by Assumptions \ref{as:2.2}.a, \ref{as:2.2}.b and \ref{as:2.4}. Further, as $\hat{\theta}(\X_{1:T_E})\overset{p}{\to} \theta_0\in\mathscr{V}(\theta_0)$ and $\mathscr{V}(\theta_0)$ is open, we have  $\PP\big[ \ddot{\theta}_T \notin \mathscr{V}(\theta_0)\big]\to 0$ and $\hat{R}_T^{SPL}=o_p(1)$ follows. Moreover, $w_T^{SPL}$ is $\mathcal{I}_{1:T}$-measurable and $O_p(1)$ by Assumptions \ref{as:2.0}.c and \ref{as:2.0}.e and $\int h \:d\bar{\PP}\big[\hat{Z}_T^{SPL}\leq \cdot|\mathcal{I}_{1:T}\big]=\int h \:d\widehat{G}_T (\cdot) \overset{p}{\to}\int h \: dG_{\infty}$ for each $h \in BL$. Replacing $Z_T$, $R_T$, $w_T$ and $\mathcal{I}_T$ in Lemma \ref{lem:2.4} by $\hat{Z}_T^{SPL}$, $\hat{R}_T^{SPL}$, $w_T^{SPL}$ and $\mathcal{I}_{1:T}$ implies that $\widehat{F_T^{SPL}}(\cdot)=\bar{\PP}\big[w_T^{SPL}\hat{Z}_T^{SPL}+\hat{R}_T^{SPL}\leq \cdot\big|\mathcal{I}_{1:T}\big]$ and $\bar{\PP}[w_T^{SPL}Z_{\infty} \leq \cdot|\mathcal{I}_{1:T}]=F_{\infty,T} ^{SPL}(\cdot|\mathcal{I}_{T_p:T})$ merge in probability.

The proof of statement 2(b) is similar to the proof of claim 1(b). As $F_T^{SPL}(\cdot|\mathcal{I}_{T_P:T})$ and $\widehat{F_T^{SPL}}(\cdot)$ merge in probability and $\widehat{F_T^{SPL}}(\cdot)$ is assumed to be stochastically uniformly continuous, Lemma \ref{lem:2.7} applies. Replacing
 $F_T$, $G_T$, $M_T$ and $\mathcal{I}_T$ by  $F_T^{SPL}(\cdot|\mathcal{I}_{T_P:T})$, $\widehat{F_T^{SPL}}(\cdot)$, $m_T\big(\hat{\psi}_{T+1}^{SPL}- \psi_{T+1} \big)$ and $\mathcal{I}_{T_P:T}$, respectively, it follows that
\begin{align*}
\quad &\PP \Big[  I_\gamma^{SPL}(\X_{T_P:T},\X_{1:T_E}) \ni \psi_{T+1} \Big|\mathcal{I}_{T_P:T}\Big]\\
&=\PP \Big[ \widehat{F_T^{SPL}}^{-1}(\gamma_1)\leq m_T\big(\hat{\psi}_{T+1}^{SPL}- \psi_{n+1} \big) \leq \widehat{F_T^{SPL}}^{-1}(1-\gamma_2) \Big|\mathcal{I}_{T_P:T}\Big] \overset{p}{\to}1-\gamma\:.\qquad \qed
\end{align*}

\bigskip
\noindent
\textbf{Proof of Theorem \ref{thm:2.3}.}
Consider the first statement and expand $\hat\psi_{T+1} (\X_{1:T}, \X_{1:T}) - \hat\psi_{T+1} (\X_{T_P:T}^c, \X_{1:T_E})= \big(\hat\psi_{T+1} (\X_{1:T}, \X_{1:T}) - \psi_{T+1}\big) -\big( \hat\psi_{T+1} (\X_{T_P:T}^c, \X_{1:T_E})-\psi_{T+1}\big) $.
We show that both terms are $o_p(1)$. Using \eqref{eq:2.4.5}, we have
\begin{align*}
\hat\psi_{T+1} (\X_{T_P:T}^c, \X_{1:T_E}) - \psi_{T+1}&= \frac{\partial \psi_{T+1}^s (\X_{T_P:T}^c;\theta_0)}{\partial \theta'} \big(\hat{\theta}(\X_{1:T_E}) - \theta_0\big) + m_T^{-1} R_T^{SPL},
\end{align*}
where $\frac{\partial \psi_{T+1}^s (\X_{T_P:T}^c;\theta_0)}{\partial \theta'}=O_p(1)$ by Assumptions \ref{as:2.0}.c and \ref{as:2.0}.e and $\hat{\theta}(\X_{1:T_E}) - \theta_0=o_p(1)$ by Assumptions \ref{as:2.0}.a, \ref{as:2.2}.a and \ref{as:2.2}.b. Together with $R_T^{SPL}=o_p(1)$ by Lemma \ref{lem:2.1} and $m_T^{-1}=o(1)$, it implies that $\hat\psi_{T+1} (\X_{T_P:T}^c, \X_{1:T_E})-\psi_{T+1}=o_p(1)$. In addition, replacing $\mathbf{Y}_{1:T}$ by $\mathbf{X}_{1:T}$ in equation \eqref{eq:2.4.4}, we get
\begin{align*}
\hat\psi_{T+1} (\X_{1:T}, \X_{1:T})- \psi_{T+1} &=  \frac{\partial \psi_{T+1}^s(\X_{1:T}, \theta_0)}{\partial \theta'}  \big(\hat{\theta}(\X_{1:T}) - \theta_0\big) +m_T^{-1} R_T^{STA},
\end{align*}
where $R_T^{STA}$ is obtained by replacing $\mathbf{Y}_{1:T}$ by $\mathbf{X}_{1:T}$ in $R_T^{2IP}$.
We have
$\Big|\Big|\frac{\partial \psi_{T+1}^s(\X_{1:T}, \theta_0)}{\partial \theta}\Big|\Big|=O_p(1)$ by Assumptions \ref{as:2.0}.c and \ref{as:2.0}.e and $\hat{\theta}(\mathbf{X}_{1:T}) - \theta_0=o_p(1)$ by Assumption \ref{as:2.0}.a. Since  $R_T^{2IP}=o_p(1)$ has been shown in Lemma \ref{lem:2.1} without using Assumption \ref{as:2.1}.b, we have $R_T^{STA}=o_p(1)$. Together with $m_T^{-1}=o(1)$, it follows that $\hat\psi_{T+1} (\X_{1:T}, \X_{1:T})-\psi_{T+1}=o_p(1)$ completing the claim.

Consider the second statement and let $\hat{Z}_T^{STA}$ follow the mixture distribution $\widehat{G}_T (\cdot)$ as a function of $\X_{1:T}$ such that given $\X_{1:T}$ the conditional cdf of the random variable $\hat{Z}_T^{STA}$ is $\widehat{G}_T (\cdot)$. Further, let 
\begin{align}
\label{eq:2.A.5}
\mbox{$\widehat{F_T^{STA}}(\cdot)$ be the conditional cdf of $\hat{w}_T^{STA}\hat{Z}_T^{STA}$ given $\mathcal{I}_{1:T}$}
\end{align}
where $\hat{w}_T^{STA}$ equals the transpose of $\frac{\partial \psi_T^s(\mathbf{X}_{1:T};\hat{\theta}(\mathbf{X}_{1:T}))}{\partial \theta}$. First, we show that $\widehat{F_T^{STA}}(\cdot)$ and $\widehat{F_T^{SPL}}(\cdot)$, defined in \eqref{eq:2.A.4},  merge in probability. The triangle inequality implies
\begin{align*}
&d_{BL}\Big(\widehat{F_T^{SPL}}(\cdot),\widehat{F_T^{STA}}(\cdot)\Big)
\leq d_{BL}\Big(\widehat{F_T^{SPL}}(\cdot),F_T^{SPL}(\cdot|\mathcal{I}_{T_P:T})\Big)\\
&\qquad+ d_{BL}\Big(F_T^{SPL}(\cdot|\mathcal{I}_{T_P:T}),F_T^{2IP}(\cdot|\mathcal{I}_{1:T})\Big) +d_{BL}\Big(F_T^{2IP}(\cdot|\mathcal{I}_{1:T}),\widehat{F_T^{STA}}(\cdot)\Big)\:,
\end{align*}
where the first two terms on the right hand side converge in probability to zero by Theorem 2.2(a) and Theorem 1, respectively. We are left to show that $F_T^{2IP}(\cdot|\mathcal{I}_{1:T})$ and $\widehat{F_T^{STA}}(\cdot)$ merge in probability. The triangle inequality implies that
\begin{align*}
&d_{BL}\Big(F_T^{2IP}(\cdot|\mathcal{I}_{1:T}),\widehat{F_T^{STA}}(\cdot)\Big)
\leq d_{BL}\Big(F_T^{2IP}(\cdot|\mathcal{I}_{1:T}),F_{\infty,T}^{2IP}(\cdot|\mathcal{I}_{1:T})\Big)+d_{BL}\Big(F_{\infty,T}^{2IP}(\cdot|\mathcal{I}_{1:T}),\widehat{F_T^{STA}}(\cdot)\Big),
\end{align*}
where $F_{\infty,T}^{2IP}(\cdot|\mathcal{I}_{1:T})$ is defined in equation \eqref{eq:2.A.1}. In the proof of Theorem \ref{thm:2.1}, we have shown that $F_T^{2IP}(\cdot|\mathcal{I}_{1:T})$ and $F_{\infty,T}^{2IP}(\cdot|\mathcal{I}_{1:T})$ merge in probability under Assumptions \ref{as:2.0} and \ref{as:2.1}. It suffices to show that $F_{\infty,T}^{2IP}(\cdot|\mathcal{I}_{1:T})$ and $\widehat{F_T^{STA}}(\cdot)$ merge in probability. Write $\hat{w}_T^{STA}\hat{Z}_T^{STA}=w_T^{2IP}\hat{Z}_T^{STA}+\hat{R}_T^{STA}$ with $\hat{R}_T^{STA}= (\hat{w}_T^{STA}-w_T^{2IP})\hat{Z}_T^{STA}$ (note that, in contrast to $\hat{w}_T^{STA}$, there is no need to introduce $w_T^{STA}$ as it equals $w_T^{2IP}$). First, we show $\hat{R}_T^{STA}=o_p(1)$. Take an arbitrary $\epsilon>0$. We obtain
\begin{align*}
&\PP \Big[\big|\hat{R}_{T}^{STA}\big|\geq\epsilon\Big]
\leq \PP \Bigg[\bigg|\bigg|\frac{\partial^2 \psi_{T+1}^s(\mathbf{X}_{1:T};\dddot{\theta}_T)}{\partial \theta \partial \theta'}\bigg|\bigg| \:\Big|\Big| \hat{\theta}(\X_{1:T}) - \theta_0\Big|\Big|\:\big|\big|\hat{Z}_T^{STA}\big|\big|\geq\epsilon\Bigg]\\
&\leq \PP \Bigg[\bigg|\bigg|\frac{\partial^2 \psi_{T+1}^s(\mathbf{X}_{1:T};\dddot{\theta}_T)}{\partial \theta \partial \theta'}\bigg|\bigg| \:\Big|\Big| \hat{\theta}(\X_{1:T}) - \theta_0\Big|\Big|\: \big|\big|\hat{Z}_T^{STA}\big|\big|\geq\epsilon \bigcap \dddot{\theta}_T \in \mathscr{V}(\theta_0)\Bigg] + \PP\Big[ \dddot{\theta}_T \notin \mathscr{V}(\theta_0)\Big]\\
&\leq \PP \Bigg[\sup_{\theta \in  \mathscr{V}(\theta_0)}\bigg|\bigg|\frac{\partial^2 \psi_{T+1}^s(\mathbf{X}_{1:T};\theta)}{\partial \theta \partial \theta'}\bigg|\bigg| \:\Big|\Big| \hat{\theta}(\X_{1:T}) - \theta_0\Big|\Big|\: \big|\big|\hat{Z}_T^{STA}\big|\big|\geq\epsilon \Bigg] + \PP\Big[ \dddot{\theta}_T \notin \mathscr{V}(\theta_0) \Big]\:.
\end{align*}
where $\dddot{\theta}_T$ lies between $\hat{\theta}(\X_{1:T})$ and $\theta_0$. The first term vanishes since $\sup\limits_{\theta \in \mathscr{V}(\theta_0)}\Big|\Big|\frac{\partial^2 \psi_{T+1}^s(\mathbf{X}_{1:T};\theta)}{\partial \theta \partial \theta'}\Big|\Big|$ is $O_{p}(1)$ by Assumptions \ref{as:2.0}.d and \ref{as:2.0}.e,  $\big|\big|\hat{\theta}(\X_{1:T}) - \theta_0\big|\big|=O_p(m_T^{-1})$ by Assumption \ref{as:2.0}.a and $\big|\big|\hat{Z}_T^{STA}\big|\big|=O_p(1)$ as $\hat{Z}_T^{STA}\sim \widehat{G}_T (\cdot) \overset{p}{\to}G_{\infty}$ by Assumption \ref{as:2.4}. Further, as $\hat{\theta}(\X_{1:T})\overset{p}{\to} \theta_0\in\mathscr{V}(\theta_0)$ and $\mathscr{V}(\theta_0)$ is open, we have  $\PP\big[ \dddot{\theta}_T \notin \mathscr{V}(\theta_0)\big]\to 0$ and $\hat{R}_T^{STA}=o_p(1)$ follows. Moreover, $w_T^{2IP}$ is $\mathcal{I}_{1:T}$-measurable and $O_p(1)$ by Assumptions \ref{as:2.0}.c and \ref{as:2.0}.e and $\int h \:d\bar{\PP}\big[\hat{Z}_T^{STA}\leq \cdot|\mathcal{I}_{1:T}\big]=\int h \:d \widehat{G}_T (\cdot) \overset{p}{\to}\int h \: dG_{\infty}$ for each $h \in BL$. Replacing $Z_T$, $R_T$, $w_T$ and $\mathcal{I}_T$ in Lemma \ref{lem:2.4} by $\hat{Z}_T^{STA}$, $\hat{R}_T^{STA}$, $w_T^{2IP}$ and $\mathcal{I}_{1:T}$ implies that $\widehat{F_T^{STA}}(\cdot)=\bar{\PP}[w_T^{2IP}\hat{Z}_T^{STA}+\hat{R}_T^{STA}\leq \cdot|\mathcal{I}_{1:T}]$ and $\bar{\PP}[w_T^{2IP}Z_{\infty} \leq \cdot|\mathcal{I}_{1:T}]=F_{\infty,T}^{2IP}(\cdot|\mathcal{I}_{1:T})$ merge in probability. Thus, $\widehat{F_T^{SPL}}(\cdot)$ and $\widehat{F_T^{STA}}(\cdot)$ merge in probability. Together with $\widehat{F_T^{SPL}}(\cdot)$ being stochastically pointwise continuous at $\gamma_1$ and $1-\gamma_2$, assertion \eqref{eq:2.4.13} follows by Lemma \ref{lem:2.8}, which completes the proof. \qed

\section{Additional Proofs}
\label{app:B}

\subsection{Proofs of Lemmas}
\label{app:B.1}

\textbf{Proof of Lemma 1.}
Consider (i). By Assumption 1.b one can write $R_T^{2IP}$ as follows:
\begin{align*}
	R_T^{2IP}=&\underbrace{m_T \big(\psi_{T+1}^s(\mathbf{X}_{1:T};\theta_0)-\psi_{T+1}(X_T,X_{T-1},\dots;\theta_0)\big)}_{=R_{1,T}^{2IP}}\\
 &\qquad   +\underbrace{\big(\hat{\theta}(\mathbf{Y}_{1:T}) - \theta_0\big)'\frac{\partial^2 \psi_{T+1}^s(\mathbf{X}_{1:T};\dot{\theta}_T)}{\partial \theta \partial \theta'} m_T \big(\hat{\theta}(\mathbf{Y}_{1:T}) - \theta_0\big)}_{=R_{2,T}^{2IP}}\:,
\end{align*}
where $\dot{\theta}_T$ lies between $\theta_0$ and $\hat{\theta}(\mathbf{Y}_{1:T})$. By Assumption 1.e, $R_{1,T}^{2IP}$ is $o_{p}(1)$; hence we are left to show that $R_{2,T}^{2IP}=o_{p}(1)$. Take an arbitrary $\epsilon>0$. We obtain
\begin{align*}
&\PP \Big[\big|R_{2,T}^{2IP}\big|\geq\epsilon\Big]
\leq \PP \Bigg[\bigg|\bigg|\frac{\partial^2 \psi_{T+1}^s(\mathbf{X}_{1:T};\dot{\theta}_T)}{\partial \theta \partial \theta'}\bigg|\bigg| \:m_T\Big|\Big| \hat{\theta}(\mathbf{Y}_{1:T}) - \theta_0\Big|\Big|^2\geq\epsilon\Bigg]\\
&\leq \PP \Bigg[\bigg|\bigg|\frac{\partial^2 \psi_{T+1}^s(\mathbf{X}_{1:T};\dot{\theta}_T)}{\partial \theta \partial \theta'}\bigg|\bigg| \:m_T\Big|\Big| \hat{\theta}(\mathbf{Y}_{1:T}) - \theta_0\Big|\Big|^2\geq\epsilon \bigcap \dot{\theta}_T \in \mathscr{V}(\theta_0)\Bigg] + \PP\Big[ \dot{\theta}_T \notin \mathscr{V}(\theta_0)\Big]\\
&\leq \PP \Bigg[\sup_{\theta \in  \mathscr{V}(\theta_0)}\bigg|\bigg|\frac{\partial^2 \psi_{T+1}^s(\mathbf{X}_{1:T};\theta)}{\partial \theta \partial \theta'}\bigg|\bigg| \:m_T\Big|\Big| \hat{\theta}(\mathbf{Y}_{1:T}) - \theta_0\Big|\Big|^2\geq\epsilon \Bigg] + \PP\Big[ \dot{\theta}_T \notin \mathscr{V}(\theta_0) \Big]\:.
\end{align*}
The first term vanishes since $\sup_{\theta \in  \mathscr{V}(\theta_0)}\Big|\Big|\frac{\partial^2 \psi_{T+1}^s(\mathbf{X}_{1:T};\theta)}{\partial \theta \partial \theta'}\Big|\Big|=O_{p}(1)$ by Assumptions 1.d and 1.e and  $m_T\big|\big|\hat{\theta}(\mathbf{Y}_{1:T}) - \theta_0\big|\big|^2=O_p(m_T^{-1})$ by Assumptions 1.a and 2.a. Further, as $\hat{\theta}(\mathbf{Y}_{1:T})\overset{p}{\to} \theta_0\in\mathscr{V}(\theta_0)$ and $\mathscr{V}(\theta_0)$ is open, we have  $\PP\big[ \dot{\theta}_T \notin \mathscr{V}(\theta_0)\big]\to 0$ and $R_{2,T}^{2IP}=o_{p}(1)$ follows.

The proof of (ii) is analogous; by Assumption 1.b one can express $R_T^{SPL}$ as follows:
\begin{align*}
    R_T^{SPL}=&\underbrace{m_T \big(\psi_{T+1}^s(\mathbf{X}_{T_P:T}^c;\theta_0)-\psi_{T+1}(X_T,X_{T-1},\dots;\theta_0)\big)}_{=R_{1,T}^{SPL}}\\
    &\qquad +\underbrace{\big(\hat{\theta}(\mathbf{X}_{1:T_E}) - \theta_0\big)'\frac{\partial^2 \psi_{T+1}^s(\mathbf{X}_{T_P:T}^c;\ddot{\theta}_T)}{\partial \theta \partial \theta'} m_T \big(\hat{\theta}(\X_{1:T_E}) - \theta_0\big)}_{=R_{2,T}^{SPL}}
\end{align*}
with $R_{1,T}^{SPL}=o_{p}(1)$ by Assumption \ref{as:2.0}.e and $\ddot{\theta}_T$ lying between $\theta_0$ and $\hat{\theta}(\mathbf{X}_{1:T_E})$. For an arbitrary $\epsilon>0$, we obtain
\begin{align*}
&\PP \Big[\big|R_{2,T}^{SPL}\big|\geq\epsilon\Big]
\leq \PP \Bigg[\bigg|\bigg|\frac{\partial^2 \psi_{T+1}^s(\mathbf{X}_{T_P:T}^c;\ddot{\theta}_T)}{\partial \theta \partial \theta'}\bigg|\bigg| \:m_T\Big|\Big| \hat{\theta}(\X_{1:T_E}) - \theta_0\Big|\Big|^2\geq\epsilon\Bigg]\\
&\leq \PP \Bigg[\bigg|\bigg|\frac{\partial^2 \psi_{T+1}^s(\mathbf{X}_{T_P:T}^c;\ddot{\theta}_T)}{\partial \theta \partial \theta'}\bigg|\bigg| \:m_T\Big|\Big| \hat{\theta}(\X_{1:T_E}) - \theta_0\Big|\Big|^2\geq\epsilon \bigcap \ddot{\theta}_T \in \mathscr{V}(\theta_0)\Bigg] + \PP\Big[ \ddot{\theta}_T \notin \mathscr{V}(\theta_0)\Big]\\
&\leq \PP \Bigg[\sup_{\theta \in  \mathscr{V}(\theta_0)}\bigg|\bigg|\frac{\partial^2 \psi_{T+1}^s(\mathbf{X}_{T_P:T}^c;\theta)}{\partial \theta \partial \theta'}\bigg|\bigg| m_T\Big|\Big| \hat{\theta}(\X_{1:T_E}) - \theta_0\Big|\Big|^2\geq\epsilon \Bigg] + \PP\Big[ \ddot{\theta}_T \notin  \mathscr{V}(\theta_0)\Big]\:.
\end{align*}
The first term vanishes as $\sup\limits_{\theta \in  \mathscr{V}(\theta_0)}\Big|\Big|\frac{\partial^2 \psi_{T+1}^s(\mathbf{X}_{T_P:T}^c;\theta)}{\partial \theta \partial \theta'}\Big|\Big|=O_{p}(1)$ by Assumptions \ref{as:2.0}.d and \ref{as:2.0}.e and  $m_T\big|\big| \hat{\theta}(\X_{1:T_E}) - \theta_0\big|\big|^2=O_p(m_T^{-1})$ by Assumptions \ref{as:2.0}.a and \ref{as:2.2}.a. Further, as $\hat{\theta}(\X_{1:T_E})\overset{p}{\to} \theta_0\in\mathscr{V}(\theta_0)$ and $\mathscr{V}(\theta_0)$ is open, we have $\PP\big[ \ddot{\theta}_T \notin \mathscr{V}(\theta_0)\big]\to 0$ and $R_{2,T}^{SPL}=o_{p}(1)$ follows.\qed \\ \\
\noindent
\textbf{Proof of Lemma 2.}
Consider (i) and let $G_{T}^{2IP}$ denote the unconditional distribution of $m_{T}\big(\hat{\theta}(\mathbf{Y}_{1:T}) - \theta_0\big)$. By Assumption \ref{as:2.1}.b, we have for each $h \in BL$
\begin{align*}
\int h \: dG_{T}^{2IP}(\cdot|\mathcal{I}_{1:T})=\int h \: dG_{T}^{2IP}\to \int h \: dG_{\infty},
\end{align*}
where the last assertion comes from Assumptions \ref{as:2.0}.a and \ref{as:2.1}.a and Portmanteau's Lemma (cf.\ \citeauthor{van2000asymptotic}, \citeyear{van2000asymptotic}; Lem.\ 2.2). Consider (ii); for each $h \in BL$ we obtain
\begin{align*}
\int h \: d\big( G_{T_E}^{SPL}(\cdot|\mathcal{I}_{T_P:T})-G_{\infty}\big)=\underbrace{\int h \: d\big( G_{T_E}^{SPL}-G_{\infty}\big)}_{I}+\underbrace{\int h \: d\big( G_{T_E}^{SPL}(\cdot|\mathcal{I}_{T_P:T})-G_{T_E}^{SPL}\big)}_{II},
\end{align*}
where $I\to 0$ by Assumptions \ref{as:2.0}.a, \ref{as:2.2}.a and \ref{as:2.2}.b and Portmanteau's Lemma and $II\overset{p}{\to}0$  by Assumption \ref{as:2.2}.c. \qed \\ \\
\noindent
\textbf{Proof of Lemma 3.}
For $r=1$ Lemma \ref{lem:2.3} appears as Lemma 2 of the supplemental material to \cite{CastilloAoS15}. Extending their result to $r>1$ we closely follow 
the proof of \citeauthor{dudley2002real} (\citeyear{dudley2002real}, Thm.\ 11.3.3) and write $Q_T$ and $Q$ to denote the probability measures corresponding to $G_T$ and $G$, respectively. Let $\epsilon>0$ and take a compact set $K\subset \R^r$ such that $Q(K)>1-\epsilon$. The set of functions $h\in \mathscr{H}$, restricted to $K$, form a compact set of functions for the supremum norm by the Arzela-Ascoli theorem (cf.\ \citeauthor{dudley2002real}, \citeyear{dudley2002real}, Thm.\ 2.4.7). Thus for some finite $J=J(\epsilon)$ there are $h_1,\dots,h_J \in \mathscr{H}$ such that for any $h \in \mathscr{H}$, there is a $j\leq J$ with $\sup_{y \in K}\big|h(y)-h_j(y)\big|<\epsilon$. Let $K^\epsilon=\{y \in \R^r:||x-y||<\epsilon \text{ for some } x \in K\}$. One has $\sup_{x \in K^\epsilon}\big|h(x)-h_j(x)\big|<3\epsilon$, since if $y \in K$ and $||x-y||<\epsilon$, then
\begin{align*}
\big|h(x)-h_j(x)\big|\leq& \big|h(x)-h(y)\big|+\big|h(y)-h_j(y)\big|+\big|h_j(y)-h_j(x)\big|\\
\leq& ||h||_{BL}||x-y||+\epsilon+||h_j||_{BL}||x-y||<3\epsilon\:.
\end{align*}
Let $g(x)=\max\{0,1-||x-K||/\epsilon\}$, where $||x-K||=\inf\{||x-y||:y\in K\}$ for all $x \in \R^r$. Then $g \in BL$ and $I\{x \in K\}\leq g \leq I\{x \in K^\epsilon\}$, where $I\{\cdot\}$ denotes the indicator function. It follows that
\begin{align*}
Q_T(\R^r \setminus K^\epsilon)=1-Q_T(K^\epsilon)\leq  1 -\int g\:dQ_T \overset{p}{\to}1- \int g \:dQ \leq 1 - Q(K)<\epsilon
\end{align*}
or equivalently $\PP\big[Q_T(\R^r\setminus K^\epsilon)\geq\epsilon\big]\to 0$. Thus, for each $h \in \mathscr{H}$ and $h_j$ as above
\begin{align*}
\sup_{h \in \mathscr{H}}\bigg|\int h\:d(Q_T-Q)\bigg|\leq& \sup_{h \in \mathscr{H}}\int \big|h-h_j\big|\:d(Q_T+Q)+\bigg|\int h_j\:d(Q_T-Q)\bigg|\\
\leq & 2(Q_T+Q)(\R^r \setminus K^\epsilon)+6\epsilon+\max_{1\leq j\leq J}\bigg|\int h_j\:d(Q_T-Q)\bigg|\\
\leq & 8\epsilon+ 2Q_T(\R^r\setminus K^\epsilon)+\max_{1\leq j\leq J}\bigg|\int h_j\:d(Q_T-Q)\bigg|\:.
\end{align*}
Hence,
\begin{align*}
&\PP\bigg[\sup_{h \in \mathscr{H}}\bigg|\int h\:d(Q_T-Q)\bigg|\geq 11\epsilon\bigg]\\
&\leq \PP\bigg[2Q_T(\R^r \setminus K^\epsilon)+\max_{1\leq j\leq J}\bigg|\int h_j\:d(Q_T-Q)\bigg|\geq 3\epsilon\bigg]\\
&\leq \PP\bigg[Q_T(\R^r \setminus K^\epsilon)\geq \epsilon\bigg]+\PP\bigg[\max_{1\leq j\leq J}\bigg|\int h_j\:d(Q_T-Q)\bigg|\geq \epsilon\bigg]\\
&\leq \PP\bigg[Q_T(\R^r\setminus K^\epsilon)\geq \epsilon\bigg]+\sum_{j=1}^J\PP\bigg[\bigg|\int h_j\:d(Q_T-Q)\bigg|\geq \epsilon\bigg]\:,
\end{align*}
where the last two terms are converging to $0$ for finite $J$ noting that $\int h_j\:d(Q_T-Q)=\int h_j\:d(G_T-G)\overset{p}{\to}0$. Observing that $\sup\limits_{h \in \mathscr{H}}\big|\int h\:d(Q_T-Q)\big|=\sup\limits_{h \in \mathscr{H}}\big|\int h\:d(G_T-G)\big|$ completes the proof.  \qed \\ \\
\noindent
\textbf{Proof of Lemma 4.}
This lemma is related to Lemma 8 in \cite{belyaev2000weakly} where the quantity corresponding to $\bar{\PP}[w_T'Z_{\infty} \leq \cdot|\mathcal{I}_T]$ is non-random.

Set $\mathscr{F}=\big\{f:\R \to \R: ||f||_{BL}\leq 1 \big \}$. The triangle inequality implies
\begin{align*}
&\sup\limits_{f \in \mathscr{F}}\bigg|\!\int\! \big[f(w_T'Z_T+R_T) - f(w_T'Z_{\infty})\big] d\bar{\PP}[\cdot|\mathcal{I}_T]\bigg|\\
&\leq\! \underbrace{\sup\limits_{f \in \mathscr{F}}\bigg|\!\int\! \big[f(w_T'Z_T+R_T\big) - f\big(w_T'Z_T)\big] d\bar{\PP}[\cdot|\mathcal{I}_T]\bigg|}_{=I}+ \underbrace{\sup\limits_{f \in \mathscr{F}}\bigg|\!\int\! \big[f(w_T'Z_T) - f(w_T'Z_{\infty})\big] d\bar{\PP}[\cdot|\mathcal{I}_T]\bigg|}_{=II}.
\end{align*}
We show that $I\overset{p}{\to}0$ and $II\overset{p}{\to}0$. Let $\epsilon>0$; as $||f||_{BL}\leq 1$ for all $f\in \mathscr{F}$ we obtain
\begin{align*}
I&\leq
 \sup\limits_{f \in \mathscr{F}}\int \Big|f(w_T'Z_T+R_T) - f(w_T'Z_T)\Big| d\:\PP[\cdot|\mathcal{I}_T]\\
%%%
=&
\sup\limits_{f \in \mathscr{F}}\int\limits_{|R_T|\leq \epsilon} \big|f(w_T'Z_T+R_T) - f(w_T'Z_T)\big| d\:\bar{\PP}[\cdot|\mathcal{I}_T]\\
&\qquad + \sup\limits_{f \in \mathscr{F}}\int\limits_{|R_T|> \epsilon} \big|f(w_T'Z_T+R_T) - f(w_T'Z_T)\big| d\:\bar{\PP}[\cdot|\mathcal{I}_T]\\
%%%
\leq& \sup\limits_{f \in \mathscr{F}}\int\limits_{|R_T|\leq \epsilon} ||f||_{BL}\big|w_T'Z_T+R_T - w_T'Z_T\big| d\:\bar{\PP}[\cdot|\mathcal{I}_T]\\
&\qquad + \sup\limits_{f \in \mathscr{F}}\int\limits_{|R_T|> \epsilon} \Big(|f(w_T'Z_T+R_T)| + |f(w_T'Z_T)|   \Big)d\:\bar{\PP}[\cdot|\mathcal{I}_T]\\
%%%
\leq& \sup\limits_{f \in \mathscr{F}}||f||_{BL}\int\limits_{|R_T|\leq \epsilon} |R_T|\: d\:\PP[\cdot|\mathcal{I}_T]+ 2 \sup\limits_{f \in  \mathscr{F}}||f||_{BL}\int\limits_{|R_T|> \epsilon}  d\:\bar{\PP}[\cdot|\mathcal{I}_T]\\
%%%
\leq& \int\limits_{|R_T|\leq \epsilon} \epsilon\: d\:\bar{\PP}[\cdot|\mathcal{I}_T]+ 2  \:\bar{\PP}\big[|R_T|> \epsilon\big|\mathcal{I}_T\big]
%%%
\leq \epsilon+ 2  \:\bar{\PP}\big[|R_T|> \epsilon\big|\mathcal{I}_T\big]\:.
\end{align*}
In line with \citeauthor{xiong2008some} (\citeyear{xiong2008some}, Thm.\ 3.3), employing Markov's inequality we have
\begin{align*}
\bar{\PP}\big[I\geq 2\epsilon\big]\leq \bar{\PP}\Big[\bar{\PP}\big[|R_T|> \epsilon\big|\mathcal{I}_T\big]\geq\epsilon/2\Big]\leq \frac{2}{\epsilon} \bar{\PP}\big[|R_T|> \epsilon\big]\to 0
\end{align*}
as $R_T=o_p(1)$ and hence $I=o_p(1)$. Consider $II$ and let $K\geq 1$. We obtain
\begin{align*}
\bar{\PP}\big[II\geq\epsilon\big]\leq \bar{\PP}\big[ ||w_T||\geq K\big]+\bar{\PP}\big[II\geq\epsilon \cap  ||w_T||\leq K\big]\:.
\end{align*}
As $||w_T||=O_p(1)$ the first term can be made arbitrarily small by choosing $K$ large. For such $K$, consider the second term and note that
\begin{align*}
\bar{\PP}\Big[II\geq\epsilon \cap  ||w_T||\leq K\Big]=& \bar{\PP}\bigg[\sup_{f \in  \mathscr{F}}\Big|\int \big[f(w_T'Z_T) - f(w_T'Z_{\infty}) \big] d\:\bar{\PP}[\cdot|\mathcal{I}_T]\Big|\geq \epsilon \cap  ||w_T||\leq K\bigg]\\
%%%
\leq& \bar{\PP}\bigg[\sup_{g \in  \mathscr{G}}\Big|\int \big[g(Z_T) - g(Z_{\infty}) \big] d\:\bar{\PP}[\cdot|\mathcal{I}_T]\Big|\geq \epsilon \cap  ||w_T||\leq K\bigg]\\
%%%
\leq& \bar{\PP}\bigg[\sup_{g \in  \mathscr{G}}\Big|\int \big[g(Z_T) - g(Z_{\infty}) \big] d\:\bar{\PP}[\cdot|\mathcal{I}_T]\Big|\geq \epsilon\bigg]\\
%%%
=& \bar{\PP}\bigg[\sup_{g \in \mathscr{G}}\Big|\int g\: d\Big(\PP[Z_T \leq \cdot|\mathcal{I}_T]-G_{\infty}\Big)  \Big|\geq \epsilon \bigg]\:,
\end{align*}
where $\mathscr{G}=\big \{g: \R^r \rightarrow \R \big|\: g(x)=f(w'x), \mbox{ for some } f \in  \mathscr{F} \mbox{ and some } w \in \R^r \mbox{ with } ||w||\leq K \big \}$. We have that $||\cdot||_{BL}$ is uniformly bounded for $\mathscr{G}$ since for every $g \in \mathscr{G}$
\begin{align*}
||g||_{BL}=&\sup_x \big|f(w'x)\big|+ \sup_{x \neq y} \frac{\big|f(w'x)-f(w'y)\big|}{|w'x-w'y|}\frac{|w'x-w'y|}{||x-y||}\\
\leq&\sup_x \big|f(w'x)\big|+ \sup_{x \neq y} \frac{\big|f(w'x)-f(w'y)\big|}{|w'x-w'y|}||w||
\leq  ||f||_{BL}\:K \leq K\:.
\end{align*}
Thus, $||g/K||_{BL}\leq 1$ and it follows by $\bar{\PP}[Z_T \leq \cdot|\mathcal{I}_T]\overset{p}{\to}G_{\infty}$ and Lemma \ref{lem:2.3} that
\begin{align*}
\bar{\PP}\Big[II\geq\epsilon \cap  ||w_T||\leq K\Big]\leq & \bar{\PP}\bigg[\sup_{g \in \mathscr{G}}\Big|\int \frac{g}{K}\:  d\Big(\PP[Z_T \leq \cdot|\mathcal{I}_T]-G_{\infty}\Big)  \Big|\geq \frac{\epsilon}{K} \bigg]\\
\leq & \bar{\PP}\bigg[\sup_{h \in \mathscr{H}}\Big|\int h\:  d\Big(\PP[Z_T \leq \cdot|\mathcal{I}_T]-G_{\infty}\Big)  \Big|\geq \frac{\epsilon}{K} \bigg]\to 0\:,
\end{align*}
where $\mathscr{H}$ is defined in Lemma \ref{lem:2.3}. Thus, $II$ is $o_p(1)$, which completes the proof. \qed \\ \\
\noindent
\textbf{Proof of Lemma 5.}
Take $\epsilon>0$ and let $F$ and $G$ be cdfs on $\R$ with $G(\tau-\epsilon)-\epsilon \leq F(\tau)\leq G(\tau+\epsilon)+\epsilon$ for all $\tau \in \R$. Fixing $u \in(\epsilon,1-\epsilon)$, we obtain
\begin{align}
\label{eq:2.B.1}
&\inf\big\{\tau\in \R:F(\tau)\geq u+\epsilon\big\}+\epsilon \\
&\geq\inf\big\{\tau\in \R:G(\tau+\epsilon)+\epsilon\geq u+\epsilon\big\}+\epsilon \nonumber\\
&= \inf\big\{\tau\in \R:G(\tau+\epsilon)\geq u\big\}+\epsilon \nonumber\\
&=\inf\big\{\tau+\epsilon, \tau \in \R:G(\tau+\epsilon)\geq u\big\} \nonumber\\
\label{eq:2.B.2}
&=\inf\big\{\tau\in \R:G(\tau)\geq u\big\} \\
&=\inf\big\{\tau+\epsilon, \tau \in \R:G(\tau)\geq u\big\}-\epsilon \nonumber\\
&=\inf\big\{\tau\in \R:G(\tau-\epsilon)\geq u\big\}-\epsilon \nonumber\\
&=\inf\big\{\tau\in \R:G(\tau-\epsilon)-\epsilon\geq u-\epsilon\big\}-\epsilon \nonumber\\
\label{eq:2.B.3}
&\geq\inf\big\{\tau\in \R:F(\tau)\geq u-\epsilon\big\}-\epsilon \:.
\end{align}
Identifying \eqref{eq:2.B.1}, \eqref{eq:2.B.2} and \eqref{eq:2.B.3} as $F^{-1}(u+\epsilon)+\epsilon$, $G^{-1}(u)$ and $F^{-1}(u-\epsilon)-\epsilon$, respectively, completes the proof. 
\qed \\ \\
\noindent
\textbf{Proof of Lemma 6.}
Let $\epsilon,\eta >0$. As $G_T$ is stochastically uniformly equicontinuous, there exists a $\delta>0$ and an $\bar{T}_1 \in \N$ such that $\PP\Big[\sup\limits_{\tau \in \R}\sup\limits_{\tau'\in\R:|\tau-\tau'|<\delta} \big|G_T(\tau')-G_T(\tau)\big|>\epsilon\Big]<\eta $ for all $T\geq \bar{T}_1$.  Take $\varkappa=\min(\delta/2,\epsilon)$.
As $d_L(F_T,G_T)\overset{p}{\to} 0$ as $T\to \infty$, there exists an $\bar{T}_2$ such that $\PP\big[d_L(F_T,G_T)>\varkappa\big]<\eta $ for all $T\geq \bar{T}_2$. Let $\bar{T}=\max(\bar{T}_1,\bar{T}_2)$.
\begin{align*}
&\PP\bigg[\sup_{\tau \in \R}\big| F_T(\tau)-G_T(\tau)\big|>2 \epsilon \bigg]\\
&\leq \PP\bigg[\sup_{\tau \in \R}\big| F_T(\tau)-G_T(\tau)\big|>2 \epsilon \cap d_L(F_T,G_T)\leq \varkappa \bigg]+ \PP\Big[ d_L(F_T,G_T)> \varkappa \Big]\\
&\leq \PP\bigg[\varkappa+\sup_{\tau \in \R}\big| G_T(\tau\pm\varkappa)-G_T(\tau)\big|>2 \epsilon \bigg]+ \PP\Big[ d_L(F_T,G_T)> \varkappa \Big]\\
&\leq \PP\bigg[\sup_{\tau \in \R}\sup_{\tau'\in\R:|\tau-\tau'|<\delta}\big| G_T(\tau')-G_T(\tau)\big|> \epsilon \bigg]+ \PP\Big[ d_L(F_T,G_T)> \varkappa \Big]
< 2\eta
\end{align*}
for all $T\geq \bar{T}$. Since the choice of $\epsilon$ and $\eta $ was arbitrary, the desired result follows. \qed \\ \\
\noindent
\textbf{Proof of Lemma 7.}
Since $F_T$ and $G_T$ merge in probability and $d_L \leq 2 d_{BL}^{1/2}$ (cf.\ \citeauthor{huber2009robust}, \citeyear{huber2009robust}, p.\ 36; \citeauthor{dudley2002real}, \citeyear{dudley2002real}, Thm.\ 11.3.3), we have $d_L(F_T,G_T)\overset{p}{\to}0$. Let $u\in (0,1)$ and take $\epsilon>0$ sufficiently small satisfying $u \in (\epsilon,1-\epsilon)$. $\PP \big[ d_L(F_T,G_T)> \epsilon\big|\mathcal{I}_T\big]$ is $o_p(1)$ since for every $\delta>0$ the Markov inequality implies $\PP\Big[\PP \big[ d_L(F_T,G_T)> \epsilon\big|\mathcal{I}_T\big]\geq \delta\Big]\leq \frac{1}{\delta}\PP \big[  d_L(F_T,G_T)> \epsilon\big]\to 0$. Employing Lemma \ref{lem:2.5}
we derive the following bounds:
\begin{align*}
&\PP \big[ M_T \leq G_T^{-1}(u)\big|\mathcal{I}_T\big]\\
&\leq \PP \big[ M_T \leq G_T^{-1}(u) \cap d_L(F_T,G_T)\leq \epsilon\big|\mathcal{I}_T\big]+\PP \big[ d_L(F_T,G_T)> \epsilon\big|\mathcal{I}_T\big]\\
&\leq \PP \big[ M_T \leq F_T^{-1}(u+\epsilon)+\epsilon \cap d_L(F_T,G_T)\leq \epsilon\big|\mathcal{I}_T\big]+o_p(1)\\
&\leq \PP \big[ M_T \leq F_T^{-1}(u+\epsilon)+\epsilon\big|\mathcal{I}_T\big]+o_p(1)\\
&=F_T\big(F_T^{-1}(u+\epsilon)+\epsilon\big)+o_p(1)=U_T\\
\\
&\PP \big[ M_T < G_T^{-1}(u) \big|\mathcal{I}_T\big]\\
&\geq \PP \big[ M_T < G_T^{-1}(u) \cap d_L(F_T,G_T)\leq \epsilon\big|\mathcal{I}_T\big]\\
&\geq \PP \big[ M_T < F_T^{-1}(u-\epsilon)-\epsilon \cap d_L(F_T,G_T)\leq \epsilon\big|\mathcal{I}_T\big]\\
&\geq \PP \big[ M_T < F_T^{-1}(u-\epsilon)-\epsilon\big|\mathcal{I}_T\big]-\PP \big[ d_L(F_T,G_T)> \epsilon\big|\mathcal{I}_T\big]\\
&=F_T\big(F_T^{-1}(u-\epsilon)-\epsilon-\big)-o_p(1)=L_T\:,
\end{align*}
where $F_T(\cdot\:-)$ denotes the left limit of $F_T(\cdot)$. We show that $L_T$ and $U_T$ converge in probability to $u$. Regarding the lower bound $L_T$ we have
\begin{align*}
&\Big|F_T\big(F_T^{-1}(u-\epsilon)-\epsilon-\big)-u\Big|\\
&\leq \Big|F_T\big(F_T^{-1}(u-\epsilon)-\epsilon-\big)-F_T\big(F_T^{-1}(u-\epsilon)-\big)\Big|+ \Big|F_T\big(F_T^{-1}(u-\epsilon)-\big)-(u-\epsilon)\Big|+\epsilon\\
&\leq \sup_{\tau \in \R}\Big|F_T(\tau-\epsilon-)-F_T(\tau-)\Big|+\Big|F_T\big(F_T^{-1}(u-\epsilon)-\big)-(u-\epsilon)\Big|+\epsilon\\
&\leq \sup_{\tau \in \R}\Big|F_T(\tau-\epsilon-)-F_T(\tau-)\Big|+\sup_{\tau \in \R}\Big|F_T(\tau)-F_T(\tau-)\Big|+\epsilon\\
&\leq 4d_{K}(F_T,G_T)+\sup_{\tau \in \R}\Big|G_T(\tau-\epsilon-)-G_T(\tau-)\Big|+\sup_{\tau \in \R}\Big|G_T(\tau)-G_T(\tau-)\Big|+\epsilon\:,
\end{align*}
where the third inequality is due to \citeauthor{cavaliere2013wild}~(\citeyear{cavaliere2013wild}, p.\ 217).  As $d_L(F_T,G_T)\overset{p}{\to}0$ and $G_T$ is stochastically uniformly equicontinuous, Lemma \ref{lem:2.6} implies $d_{K}(F_T,G_T)\overset{p}{\to}0$. Further,  $\sup\limits_{\tau \in \R}\big|G_T\big(\tau-\epsilon-\big)-G_T\big(\tau-\big)\big|=o_p(1)$ and $\sup\limits_{\tau \in \R}\big|G_T(\tau)-G_T(\tau-)\big|=o_p(1)$ by stochastic uniform equicontinuity completing $L_T\overset{p}{\to}u$.  Regarding the upper bound $U_T$ we have
\begin{align*}
&\Big|F_T\big(F_T^{-1}(u+\epsilon)+\epsilon\big)-u\Big|\leq \Big|F_T\big(F_T^{-1}(u+\epsilon)+\epsilon\big)-F_T\big(F_T^{-1}(u+\epsilon)\big)\Big|\\ 
&\qquad+\Big|F_T\big(F_T^{-1}(u+\epsilon)\big)-F_T\big(F_T^{-1}(u+\epsilon)-\big)\Big| + \Big|F_T\big(F_T^{-1}(u+\epsilon)-\big)-(u+\epsilon)\Big|+\epsilon\\
&\leq \sup_{\tau \in \R}\Big|F_T(\tau+\epsilon)-F_T(\tau)\Big|+2\sup_{\tau \in \R}\Big|F_T(\tau)-F_T(\tau-)\Big|+\epsilon\\
&\leq 6d_{K}(F_T,G_T) +\sup_{\tau \in \R}\Big|G_T(\tau+\epsilon)-G_T(\tau)\Big|+2\sup_{\tau \in \R}\Big|G_T(\tau)-G_T(\tau-)\Big|+\epsilon\:,
\end{align*}
where all terms on the right hand side are $o_p(1)$ such that $U_T\overset{p}{\to}u$. We obtain
\begin{align*}
\underbrace{L_T(u)}_{\overset{p}{\to}u}\leq \PP \big[ M_T < G_T^{-1}(u) \big|\mathcal{I}_T\big]\leq\PP \big[ M_T \leq G_T^{-1}(u) \big|\mathcal{I}_T\big]\leq \underbrace{U_T(u)}_{\overset{p}{\to}u}\:,
\end{align*}
which implies that $\PP \big[ M_T <G_T^{-1}(u) \big|\mathcal{I}_T\big]$ and $\PP \big[ M_T \leq G_T^{-1}(u) \big|\mathcal{I}_T\big]$ converge in probability to $u$ for arbitrary $u \in (0,1)$; in particular $\gamma_1$ and $1-\gamma_2$. It follows that
\begin{align*}
\;\;\quad&\PP\big[G_T^{-1}(\gamma_1) \leq M_T \leq G_T^{-1}(1-\gamma_2) \big|\mathcal{I}_T\big]\\
&=\PP\big[ M_T \leq G_T^{-1}(1-\gamma_2) \big|\mathcal{I}_T\big] - \PP\big[M_T < G_T^{-1}(\gamma_1)\big|\mathcal{I}_T \big]
\overset{p}{\to}1-\gamma_2-\gamma_1\:.\quad \qed
\end{align*}
\noindent
\textbf{Proof of Lemma 8.}
Let $\epsilon,\eta >0$ and set $\bar{\epsilon}=\min\{\epsilon,u,1-u\}/2$. Since $G_T^{-1}$ is pointwise equicontinuous at $u$, there exist a $\delta>0$ and an $\bar{T}_1$ such that
$\PP\big[\sup\limits_{|u-v|<\delta}\big|G_T^{-1}(v)-G_T^{-1}(u)\big|>\bar{\epsilon}\big]<\eta $
for all $T\geq \bar{T}_1$. Take $\varkappa = \min\{\delta/2,\bar{\epsilon}\}$. As $d_L(F_T,G_T)\overset{p}{\to} 0$ as $T\to \infty$, there exists an $\bar{T}_2$ such that $\PP\big[d_L(F_T,G_T)>\varkappa\big]<\eta $ for all $T\geq \bar{T}_2$.
\begin{align*}
&\PP\bigg[\big| F_T^{-1}(u)-G_T^{-1}(u)\big|>2 \epsilon \bigg]\leq \PP\bigg[\big| F_T^{-1}(u)-G_T^{-1}(u)\big|>2 \bar{\epsilon} \bigg]\\
&\leq \PP\bigg[\big| F_T^{-1}(u)-G_T^{-1}(u)\big|>2 \bar{\epsilon} \cap d_L(F_T,G_T)\leq \varkappa \bigg]+ \PP\Big[ d_L(F_T,G_T)> \varkappa \Big]\\
&\leq \PP\bigg[\varkappa+\big| G_T^{-1}(u\pm\varkappa)-G_T^{-1}(u)\big|>2 \bar{\epsilon} \bigg]+ \PP\Big[ d_L(F_T,G_T)> \varkappa \Big]\\
&\leq \PP\Big[\sup\limits_{|u-v|<\delta}\big|G_T^{-1}(v)-G_T^{-1}(u)\big|>\bar{\epsilon}\Big]+ \PP\Big[ d_L(F_T,G_T)> \varkappa \Big]< 2\eta
\end{align*}
for all $T\geq \bar{T}=\max(\bar{T}_1,\bar{T}_2)$, where the third inequality follows from Lemma 5 and $u \in (\bar{\epsilon},1-\bar{\epsilon})\subseteq(\varkappa,1-\varkappa)$. As $\epsilon$ and $\eta $ were arbitrarily chosen,  this completes the proof. \qed

\subsection{Proofs of Corollaries}
\label{app:B.2}

\textbf{Proof of Corollary 1.}
Statement 1(a) follows immediately from Theorem 2.1(a) and $\widehat{F_T^{2IP}}(\cdot)$ equals to $\Phi\big(\cdot/\sqrt{\hat{\upsilon}_T^{2IP}}\big)$. Regarding claim 1(b), it is sufficient to show that $1/\hat{\upsilon}_T^{2IP}=O_p(1)$ implies that $\Phi\big(\cdot/\sqrt{\hat{\upsilon}_T^{2IP}}\big)$ is stochastically uniformly equicontinuous by Theorem 2.1(b). Since $1/\hat{\upsilon}_T^{2IP}=O_p(1)$ by assumption, we have for all $\varkappa>0$, there exist $K=K(\varkappa)$ and $\bar{T}=\bar{T}(\varkappa)$ such that $\PP\big[1/\hat{\upsilon}_T^{2IP}>K\big]<\varkappa$ for all $T>\bar{T}$. Let $\phi$ denote the standard normal density. Taking $\delta=\frac{\epsilon}{\phi(0)\sqrt{K}}$, we obtain
\begin{align*}
&\PP\Big[\sup_{\tau\in\R} \sup_{\tau':|\tau-\tau'|<\delta} \big|\Phi\big(\tau'/ \sqrt{\hat{\upsilon}_T^{2IP}}\big)-\Phi\big(\tau/ \sqrt{\hat{\upsilon}_T^{2IP}}\big)\big|>\epsilon\Big]\\
\leq& \PP\Big[\phi(0)\delta/\sqrt{\hat{\upsilon}_T^{2IP}}>\epsilon\Big]=\PP\Big[1/\hat{\upsilon}_T^{2IP}>K\Big]<\varkappa
\end{align*}
for all $T>\bar{T}$ such that the stochastic uniform equicontinuity condition holds.

Statement 2(a) follows from Theorem 2.2(a) and $\widehat{F_T^{SPL}}(\cdot)$ equals to $\Phi\big(\cdot/\sqrt{\hat{\upsilon}_T^{SPL}}\big)$. Claim 2(b) is proven analogously to the claim of 1(b) replacing $\hat{\upsilon}_T^{2IP}$ with $\hat{\upsilon}_T^{SPL}$. \qed \\ \\
\noindent
\textbf{Proof of Corollary 2.}
In the proof of Theorem 3 we have shown that $\widehat{F_T^{SPL}}(\cdot)$ and $\widehat{F_T^{STA}}(\cdot)$ merge in probability, which simplify to $\Phi\big(\cdot/\sqrt{\hat{\upsilon}_T^{SPL}}\big)$ and $\Phi\big(\cdot/\sqrt{\hat{\upsilon}_T^{STA}}\big)$, respectively, under Assumption \ref{as:2.5}.
It remains to show that $\widehat{F_T^{SPL}}^{-1}(u)=\sqrt{\hat{\upsilon}_T^{SPL}}\Phi^{-1}(u)$ is stochastically pointwise equicontinuous at $u=\gamma_1,1-\gamma_2$. First, we show that $\hat{\upsilon}_T^{SPL}=O_p(1)$. The triangle inequality implies $\hat{\upsilon}_T^{SPL}\leq\upsilon_T^{SPL}+ \big|\hat{\upsilon}_T^{SPL}-\upsilon_T^{SPL}\big|$, where 
$\upsilon_T^{SPL}=\frac{\partial \psi_{T+1}^s(\X_{T_P:T}^c;\theta_0)}{\partial \theta'}\Upsilon_0\frac{\partial \psi_{T+1}^s(\X_{T_P:T}^c;\theta_0)}{\partial \theta}$ is $O_p(1)$ by Assumptions \ref{as:2.0}.c and \ref{as:2.0}.e. Moreover, for an arbitrary $\varepsilon>0$, we have
\begin{align*}
& \PP \Bigg[\bigg|\bigg|\frac{\partial \psi_{T+1}^s(\X_{T_P:T}^c;\theta_0)}{\partial \theta}-\frac{\partial \psi_{T+1}^s(\X_{T_P:T}^c;\hat{\theta}(\X_{1:T_E})}{\partial \theta}\bigg|\bigg|\geq\epsilon\Bigg]\\
\leq& \PP \Bigg[\bigg|\bigg|\frac{\partial^2 \psi_{T+1}^s(\X_{T_P:T}^c;\tilde{\theta}_T)}{\partial \theta \partial \theta'}\bigg|\bigg| \:\Big|\Big| \hat{\theta}(\X_{1:T_E}) - \theta_0\Big|\Big|\geq\epsilon \bigcap \tilde{\theta}_T \in \mathscr{V}(\theta_0)\Bigg] + \PP\Big[ \tilde{\theta}_T \notin \mathscr{V}(\theta_0)\Big]\\
\leq& \PP \Bigg[\sup_{\theta \in  \mathscr{V}(\theta_0)}\bigg|\bigg|\frac{\partial^2 \psi_{T+1}^s(\X_{T_P:T}^c;\theta)}{\partial \theta \partial \theta'}\bigg|\bigg| \:\Big|\Big| \hat{\theta}(\X_{1:T_E}) - \theta_0\Big|\Big|\geq\epsilon \Bigg] + \PP\Big[ \tilde{\theta}_T \notin \mathscr{V}(\theta_0)\Big]\:,
\end{align*}
where $\tilde{\theta}_T$ lies between $\hat{\theta}(\X_{1:T_E})$ and $\theta_0$. The first term vanishes as $\sup\limits_{\theta \in  \mathscr{V}(\theta_0)}\Big|\Big|\frac{\partial^2 \psi_{T+1}^s(\X_{T_P:T}^c;\theta)}{\partial \theta \partial \theta'}\Big|\Big|=O_{p}(1)$ by Assumptions \ref{as:2.0}.d and \ref{as:2.0}.e and  $\big|\big|\hat{\theta}(\X_{1:T_E}) - \theta_0\big|\big|=O_p(m_T^{-1})$ by Assumptions \ref{as:2.0}.a and \ref{as:2.1}.a. Further, since $\hat{\theta}(\X_{1:T_E})\overset{p}{\to} \theta_0\in\mathscr{V}(\theta_0)$ and $\mathscr{V}(\theta_0)$ is open, we have $\PP\big[ \tilde{\theta}_T \notin \mathscr{V}(\theta_0)\big]\to 0$ and $\Big|\Big|\frac{\partial \psi_{T+1}(\X_{T_P:T};\theta_0)}{\partial \theta}-\frac{\partial \psi_{T+1}(\X_{T_P:T};\hat{\theta}(\X_{1:T_E})}{\partial \theta}\Big|\Big|=o_{p}(1)$ follows. Together with  $\hat{\Upsilon}(\X_{1:T_E})\overset{p}{\to}\Upsilon_0$, it implies $\big|\hat{\upsilon}_T^{SPL}-\upsilon_T^{SPL}\big|=o_p(1)$ and hence $\hat{\upsilon}_T^{SPL}=O_p(1)$. Next, we show that the stochastic pointwise equicontinuity condition is satisfied. For $K>0$, we get
\begin{align*}
& \PP\big[\sqrt{\hat{\upsilon}_T^{SPL}}\sup\limits_{v:|u-v|<\delta}\big|\Phi^{-1}(u)-\Phi^{-1}(v)\big|>\epsilon \Big]\\
\leq& \PP\big[\sqrt{K}\sup\limits_{v:|u-v|<\delta}\big|\Phi^{-1}(u)-\Phi^{-1}(v)\big|>\epsilon\Big]+\PP\big[\hat{\upsilon}_T^{SPL}>K\big]\:.
\end{align*}
$K$ can be chosen such that the last term is arbitrary small for large $T$ as $\hat{\upsilon}_T^{SPL}=O_p(1)$. Given $K$, the first term is $0$ by the choice of $\delta$ and continuity of $\Phi^{-1}$. \qed

\end{document}